\documentclass[usenatbib,a4paper]{mn2e}

\pdfoutput=1 

\usepackage{hyperref}
\usepackage{amsfonts}
\usepackage{amssymb}
\usepackage{amsmath}
\usepackage{graphicx}
\usepackage[totalwidth=505pt,totalheight=680pt]{geometry}

\newcommand{\adsurl}[1]{\href{#1}{ADS}}
\providecommand{\url}[1]{\href{#1}{#1}}

\newcommand{\del}{\ensuremath{\delta}}

\newcommand{\avg}[1]{\ensuremath{\langle \,#1\, \rangle}}

\newcommand{\ds}{\Delta s}
\newcommand{\eqn}[1]{equation~\eqref{#1}}

\newcommand{\dd}{\mathrm{d}}
\newcommand{\be}{\begin{equation}}
\newcommand{\ee}{\end{equation}}

\newcommand{\PS}{\mathrm{PS}}
\newcommand{\MS}{\mathrm{up}}

\title[Markov velocities]
      {On the Markovian assumption in the excursion set approach:  
       the approximation of Markovian Velocities}

\author[M.~Musso, R.~K.~Sheth]
{Marcello Musso$^{1}$\thanks{E-mail: marcello.musso@uclouvain.be} 
 \& Ravi K.~Sheth$^{2,3}$\thanks{E-mail: shethrk@physics.upenn.edu}\\
 $^1$ CP3-IRMP, Universit\'e Catholique de Louvain, 
      2 Chemin du Cyclotron, 1348 Louvain-la-Neuve, Belgium \\
 $^2$ The Abdus Salam International Center for Theoretical Physics,
      Strada Costiera, 11, Trieste 34151, Italy\\
 $^3$ Center for Particle Cosmology, University of Pennsylvania, 
      209 S. 33rd St., Philadelphia, PA 19104, USA}

\begin{document}

\pagerange{\pageref{firstpage}--\pageref{lastpage}}

\maketitle 

\label{firstpage}

\begin{abstract}
The excursion set approach uses the statistics of the density field smoothed on a wide range of scales, to gain insight into a number of interesting processes in nonlinear structure formation, such as cluster assembly, merging and clustering. The approach treats the curve defined by the height of the overdensity fluctuation field when changing the smoothing scale as a random walk. The steps of the walks are often assumed to be uncorrelated, so that the walk heights are a Markov process, even though this assumption is known to be inaccurate for physically relevant filters. 
We develop a model in which the walk steps, rather than heights, are a Markov process, and correlations between steps arise because of nearest neighbour interactions. This model is a particular case of a general class, which we call Markov Velocity models. We show how these can approximate the walks generated by arbitrary power spectra and filters, and, unlike walks with Markov heights, provide a very good approximation to physically relevant models.
We define a Markov Velocity Monte Carlo algorithm to generate walks whose first crossing distribution is very similar to that of TopHat-smoothed $\Lambda$CDM walks. Finally, we demonstrate that Markov Velocity walks generically exhibit a simple but realistic form of assembly bias, so we expect them to be useful in the construction of more realistic merger history trees. 
\end{abstract}

\begin{keywords}
large-scale structure of Universe
\end{keywords}

\section{Introduction}
Simulations of hierarchical gravitational clustering suggest that the abundance and clustering of gravitationally bound objects in the Universe can be powerful tools for constraining the nature of the initial fluctuation field.  Since simulations are expensive, there is considerable interest in models that can provide a better understanding of how cluster abundances and clustering depend on cosmological parameters.  The excursion set approach \citep{bcek91} is perhaps the most developed of these:  motivated by the seminal work of \cite{ps74} and \cite{eps83} it provides an analytical framework which relates the statistics of gravitationally bound dark matter haloes to fluctuations in the primordial density field, and the subsequent expansion history. 

In this approach, at a given (randomly chosen) position in space one looks at the overdensity field smoothed on some scale $R$: plotting this smoothed $\delta$ as a function of (the inverse of) $R$ resembles a random trajectory, the steps of which are, in general, correlated.  The nature of the correlations depends on the smoothing filter (e.g. TopHat, Gaussian), and on the nature of the initial fluctuation field (Gaussian or non-Gaussian).  Repeating this for every position in space gives an ensemble of trajectories, each one of which starts from $\delta(R=\infty)=0$ (the Universe is homogeneous on large smoothing scales).  For each trajectory, one searches for the largest $R$ (if any) for which the value of the smoothed density field lies above some threshold value (which may itself depend on $R$), the value of which is determined by the expansion history of the background cosmology.  An object of mass $M\sim  R^3$ is then associated with that trajectory.  

If $\dd n/\dd M$ denotes the comoving number density of haloes of mass $M$, then the mass fraction in such haloes is $(M/\bar\rho)\,\dd n/\dd M$, where $\bar\rho$ is the comoving background density.  The excursion set approach assumes that this halo mass fraction equals the fraction of walks which cross the threshold (the ``barrier'') for the first time when the smoothing scale is $R$:
\begin{equation}
 f(R)\,\dd R = (M/\bar\rho)\,(\dd n/\dd M)\, \dd M.
 \label{ansatz}
\end{equation}
Although recent work has focused on the shortcomings of this ansatz \citep{ps12}, not to mention the fact that variables other than the overdensity affect halo formation \citep{bm96, smt01}, and this is not evident in the simplest version of the approach outlined above \citep[but see][]{st02,scs13}, the first crossing distribution is nevertheless expected to provide substantial insight into the dependence of $\dd n/\dd M$ on cosmological parameters.  

In practice, one works not with $f(R)$ but with $f(s)$, where 
\begin{equation}
 \label{sR}
 s(R) \equiv \int \frac{\dd k}{k}\, \frac{k^3P(k)}{2\pi^2}\,W^2(kR)
\end{equation} 
denotes the variance in the fluctuation field when smoothed on scale $R$ with a filter of shape $W$.  In hierarchical models, $s$ is a monotonic function of $R$, so $f(R)\dd R = f(s)\dd s$.  Working with $s$ has the advantage of removing most of the dependence on the shape of the power spectrum:  $P(k)$ mainly matters only through the dependence of $s$ on $R$.  

To solve the first crossing problem, we must be able to identify the fraction of trajectories for which $\delta = b(s)$ and $\delta<b(S)$ for all $S < s$.  This is straightforward only when $\delta$ is a Markov process, allowing for exact analytic results for some barriers \citep{bcek91, rks98}, and so much subsequent work has focused on this approximation.  The more general case of walks with correlated steps has only recently seen real analytic progress, which is based on the realization that much of the constraint coming from the condition $\delta<b(S)$ for all $S < s$ is encapsulated by counting walks which are crossing the barrier from below \citep{bcek91, ms12}.  That is, if one defines the velocity $v \equiv \dd\delta/\dd s$, then a good approximation to $f(s)$, valid for all power-spectra, smoothing windows and barrier shapes of current interest, can be computed from the joint probability $p(\delta,v;s)$ that a walk reaches $\delta$ at scale $s$ with velocity $v \ge \dd b/\dd s$ (so that it overtakes the barrier).  That is, 
\begin{equation}
 \label{fms}
 f(s)\simeq f_\MS(s) = p(b;s) \int_{b'}^\infty \!\!\dd v\,(v-b')\,p(v|b)\,,
\end{equation}
where $b'\equiv \dd b/\dd s$.  This simplicity also holds when the fluctuation field being smoothed is not Gaussian \citep{ms13a}.  

In this approximation, the normalized quantity 
\begin{equation}
  \gamma^2 \equiv 
  \frac{\avg{\!v\delta\!}^2}{\avg{\!v^2\!}\avg{\!\delta^2\!}} 
  = \frac{1}{4s \avg{\!v^2\!}}
\label{gamma}
\end{equation} 
plays an important role, because $p(b/\sqrt{s})$ is a normalized Gaussian with zero mean, and $p(v|b)$ is also Gaussian whose mean is $\avg{\!v|b\!} \equiv \avg{\!v\delta\!}b/s = b/2s$ and whose variance is $\avg{\!v^2\!}(1 - \gamma^2)=1/4\Gamma^2 s$, where $\Gamma^2\equiv \gamma^2/(1-\gamma^2)$.  
The additional constraint on $v$ that comes from requiring walks to cross the barrier upwards makes \eqn{fms} an essentially perfect description of the first crossing distribution down to scales of order $s\lesssim \Gamma^2b^2(s)$. 

Recently, \cite{ms13b} have shown that judicious use of the same constraint on $v$
allows an even better approximation for $f(s)$.  This approach sets 
\begin{equation}
 f(s)\simeq f_\mathrm{BS}(s)\,,
 \label{bsApprox}
\end{equation}
where $f_\mathrm{BS}$ is got by solving, via back-substitution, the expression 
\begin{equation}
  p(\delta\ge b(s);s) =
 \int_0^s \dd S\,f_\mathrm{BS}(S) \,p(\delta\ge b(s)|\mathrm{up~} S)\,,
\label{fbs}
\end{equation}
where 
\begin{equation}
 p(\delta\ge b|{\rm up}\ S) \equiv 
      \frac{\int_{B'}^\infty {\rm d}V (V-B')\,p(V|B)\,p(\delta\ge b|V,B)}
           {\int_{B'}^\infty {\rm d}V (V-B')\,p(V|B)}\,,
 \label{pbsupS}
\end{equation}
and $B$ and $V$ denote the barrier height and the velocity of the walk on scale $S$.  The new ingredient required by this approach is $p(b|B,V)$, the probability that the walk has height $b$ on $s$ when it is conditioned to have height $B$ and velocity $V$ on scale $S\le s$.  That is, in contrast to $f_\MS$, which only required knowledge of walk heights and velocities on the same scale, this one requires knowledge of correlations between two, but only two, different scales.  To turn the approximation sign in our \eqn{bsApprox} into an equality, we would have to replace $p(b|{\rm up}\ S)$ with $p(b|{\rm first}\ S)$:  the additional requirement that the walk was below $B$ on all smoothing scales larger than $V$ would introduce many more scales into the problem.  Fortunately, \cite{ms13b} have shown that, in fact, $f_\mathrm{BS}(s)$ is an essentially perfect description of $f(s)$ over all values of $s$. 

The analysis that follows provides yet another example of why, in the context of random walks, thinking about $v$ is so useful: this is because, generally speaking, the derivative of a non-Markovian diffusive process is often Markovian. Section~\ref{simple} describes a toy model that is the next more complicated model to one in which there are no correlations between steps.  We use this model to motivate a more general Markov Velocities model, which we formulate and develop in Section~\ref{general}.  Our Markovian Velocities model is particularly useful in cosmology since it provides a very good description of the first crossing distribution of walks having the full correlation structure resulting from TopHat smoothed $\Lambda$CDM power spectra: some of these applications are described in Section~\ref{sec:cond}.  A final section summarizes our results.

\section{A simple illustrative example}\label{simple}

In this section we consider the first crossing distribution of discretized walks each having $n$ infinitesimal steps, and heights $\delta_1,\dots,\delta_n$.
If the underlying process is Gaussian, the joint distribution of the steps is entirely determined by the correlation matrix
 $C_{ij}\equiv\avg{\!(\delta_i-\delta_{i-1})(\delta_j-\delta_{j-1})\!}$.  
Since
 $\delta_k = \sum_{i=1}^k(\delta_i-\delta_{i-1})$, 
the variance is
 $s_k = \sum_{i,j=1}^k C_{ij}$. 
Typically, one considers evenly spaced steps, for which $s_k = k\,\ds$.
If the steps are equally spaced in variance, then the constraint
 $2\sum_{i=1}^{k-1}C_{ik}+C_{kk} = s_k-s_{k-1} = \ds$ is always satisfied, and it simply means that in the continuum limit one always has
 $2\avg{\!\delta v\!}=\avg{\!\delta^2\!}'=1$.
The analysis so far has been general.  

\begin{figure*}
 \label{fig:Cij}
 \includegraphics[width=.49\textwidth]{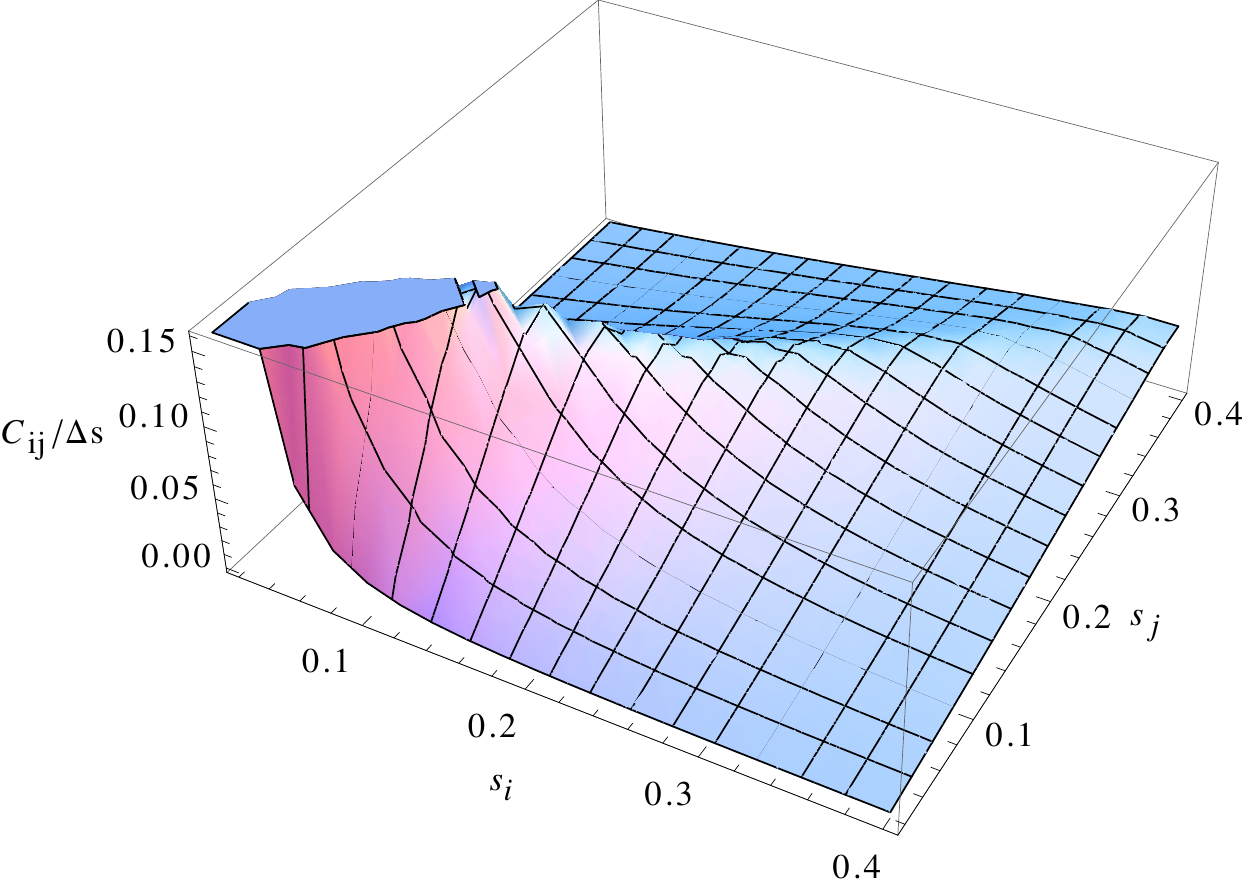}
 \includegraphics[width=.5\textwidth]{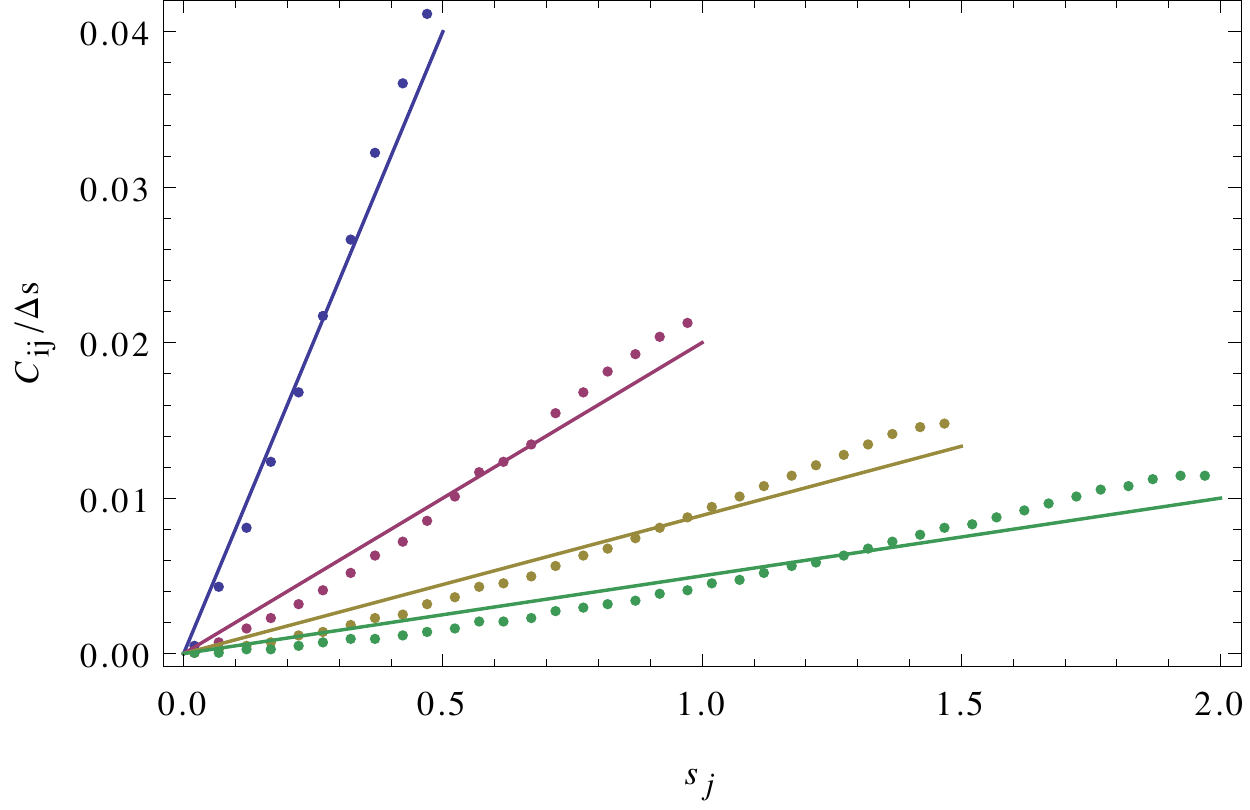}
 \caption{Left: correlation matrix $C_{ij}=\Delta s^2 \avg{v_iv_j}$ for the walk steps obtained by TopHat filtering a $\Lambda$CDM power spectrum.  The correlation is maximal along the diagonal and decreases approximately linearly with distance from the diagonal.  Right:  the same $C_{ij}$, shown for a few fixed values of $s_i = 0.5,1,1.5,2$ (symbols), and compared with the toy model $\avg{v_iv_j} = s_j/s_i^2$ described in Section~\ref{subsec:toy} (lines).}
\end{figure*}

\subsection{Markov heights from uncorrelated steps}

The usual approach to this problem assumes that $C_{ij}$ is diagonal, with  $C_{ij}=\ds\,\delta_{ij}$.  The probability $p(\delta_1,\dots,\delta_n)$ of a walk  can be equivalently expressed as the joint probability of the steps $\delta_i-\delta_{i-1}$ for $i\le n$.
The independence of the steps makes the joint probability factorize as
\begin{equation}
\label{pwalkuncorr}
  p(\del_1,\dots,\del_n) = \prod_{i=1}^n p(\delta_i-\delta_{i-1})\,,
\end{equation}
where each $\delta_i-\delta_{i-1}$ is an independent Gaussian variate with zero mean and variance $C_{ii}=\ds$.

Suppose we now consider a walk with $k$ infinitesimal steps of width $\ds$ and a finite one of width $(n-k)\ds$.  Since these are $k+1$ independent Gaussian variables, the conditional distribution of $\delta_n$ given all the other $k$ steps is a Gaussian with mean $\delta_k$ and variance $s_n-s_k$.  Since $p(\delta_n|\delta_k,\ldots,\delta_1) = p(\delta_n|\delta_k)$, the heights at all steps previous to $k$ do not matter, the walk heights are said to follow a Markov process.  It is this Markovianity which vastly simplifies analysis of the first crossing distribution.

\subsection{The next step: Markov steps}
\label{subsec:toy}

If the steps are not independent variables, as \cite{ms12} have emphasized, then in order to account for the correlations it is convenient to think in terms of the walk velocity 
 $v_i = (\delta_i-\delta_{i-1})/\ds$, 
for which the correlation matrix reads
 $C_{ij} = \ds^2\,\avg{\!v_i v_j\!}$: 
each entry is now quadratic in $\ds$.  

If $C_{ij}$ is not diagonal, the next more complicated assumption is that the correlation between the velocities is maximal for $i=j$ and decreases to a minimum for $i=0$ with constant slope:  the correlation is directly proportional to the separation between the steps. Then the constraint implies
 $C_{ij} = \ds \min(i,j)/\max^2(i,j)$. 
This condition is equivalent to having
 $\avg{\!v_i v_j\!} = s_i/s_j^2$ for $s_i\leq s_j$.
Figure~\ref{fig:Cij} shows that, despite its simplicity, a matrix with this structure is actually rather similar to that associated with TopHat smoothing of a $\Lambda$CDM power spectrum.  In the next section we use this similarity to motivate models which generalize this, our simplest toy model.  But first, we will use this simple model to illustrate a number of rather interesting features of this Markov steps model.  

First, we note that $C_{ij}$ has determinant 
 $|C|= (\ds^n/n!^4)\prod_{i=1}^n[i^3-(i-1)^3]$.  
Therefore, it is invertible, having inverse 
\begin{equation}
  C^{-1} = \frac{1}{\ds} \left[
  \begin{array}{cccc}
  x_1 & -y_2 & 0 & \ldots \\
  -y_2 & x_2 & -y_3 &  \\
  0 & -y_3 & x_3 & \ddots \\
  \vdots &  & \ddots & \ddots
  \end{array}
  \right],
\label{corrmat}
\end{equation}
where $x_i=i^4\!/[i^3-(i-1)^3] + i^4\!/[(i+1)^3-i^3]$ for all $i<n$, $x_n=n^4\!/[n^3-(n-1)^3]$ and $y_i=i^2(i-1)^2\!/[i^3-(i-1)^3]\,$.  

One can thus write the probability of the trajectory in a factorized form with only nearest-neighbour interactions:
\begin{equation}
\label{pwalk}
  p(\del_1,\dots,\del_n) = \prod_{i=1}^n p(\Lambda_i)\,,
\end{equation}
where we introduced the $n$ independent Gaussian variates 
\be
 \Lambda_i\equiv\delta_i-\delta_{i-1}-[(i-1)^2/i^2](\delta_{i-1}-\delta_{i-2}), 
\ee
whose variance is 
\be 
 \avg{\Lambda_i^2}=(\ds/i^4)[i^3-(i-1)^3].
\ee 
Note that even though $C_{ij}$ has no zero entries -- so it is very different from the case of independent steps -- the conditional distribution of one step only depends on the one immediately before, showing that this is a model where the steps rather than the heights are Markovian.

Suppose we again consider a walk with $k$ infinitesimal steps of width $\ds$ and a finite one of width $(n-k)\ds$. Since these are $k+1$ Gaussian variables, the conditional distribution of $\delta_n-\delta_k$ is the Gaussian $p(\delta_n-\delta_k-w_iC_{ij}^{-1}(\delta_j-\delta_{j-1}))$, where $w_i \equiv \avg{\!(\delta_n-\delta_k)(\delta_i-\delta_{i-1})\!}$ and $i,j\leq k$.
In this particular case, $w_i\equiv \sum_{j=k+1}^nC_{ij} = C_{ik}\sum_{j=k+1}^n(k^2/j^2)$, so that the mean still depends only on $\delta_k-\delta_{k-1}=\ds\, v_k$.  This gives
\begin{equation}
\label{condprob}
  p(\delta_n|\delta_k,\dots,\delta_1)= p(\delta_n-\delta_k-\psi_{nk} s_k v_k)
\end{equation}
with no sum running over the $k$ index, where
\begin{equation}
  \psi_{nk} \equiv
  \frac{\avg{\!(\delta_n-\delta_k)v_i\!}}{s_k\avg{\!v_kv_i\!}}
  = \!\sum_{j=k+1}^n \!\ds \,\frac{s_k}{s_j^2} \;
  \to \, \!\left(1-\frac{s_k}{s_n}\right)\,;
\label{psitoy}
\end{equation}
notice that the formal definition of $\psi_{nk}$ depends on $i$, but its explicit expression does not.
Furthermore, the last statement holds in the continuum limit, when the sum can be replaced by an integral.
The variance of the distribution is $\avg{\!(\delta_n-\delta_k)^2\!}-s_k\psi_{nk}^2$; in the continuum limit, 
since $\avg{\!(\delta_n-\delta_k)^2\!}=s_n-s_k-2\sum_{i=1}^k\sum_{j=k+1}^n \!C_{ij}$, 
it becomes $s_n(1-s_k/s_n)^3$.


For future reference, we show that the same could have been proven by noting that, after some algebra, one has
\begin{equation}
  \delta_n-\delta_k-\psi_{nk} s_k v_k =
  \sum_{i=k+1}^n \Lambda_i\sum_{j=i}^n\frac{i^2}{j^2}\,.
\label{condlargestep}
\end{equation}
This expression is explicitly independent of any of the $\Lambda_i$'s with $i\leq k$, and therefore it factorizes from $p(\delta_1,\dots,\delta_k)$.

Also for future reference, we note that our requirement that 
$\avg{\!v_i v_j\!} = s_i/s_j^2$ for $s_i\leq s_j$ means that 
$\avg{\!v_i^2\!} = 1/s_i$.  This means that $\gamma$,
the quantity which plays an important role in the analysis of $f_\MS$, equals $1/2$ for all scales, and $\Gamma^2=\gamma^2/(1-\gamma^2)=1/3$.  
  
Finally, note that the term on the right-hand side of \eqn{condprob} is the same object which plays a key role in \eqn{pbsupS} of the back-substitution approach.  Since $p(b|B,V)$ here is independent of what happened on scales larger than $S$, it is in fact the same as $p(b|{\rm first}\ S)$, which means that \eqn{bsApprox} should be an even better approximation for this toy model than it is in general.  

\subsection{Monte Carlo realizations}
The factorized form of the joint probability (equation~\ref{pwalk}) makes the model introduced above particularly well suited for numerical simulation.  This is because, as can be seen by setting $k=0$ in the previous equation, the walk height after $n$ steps is 
\begin{equation}
 \label{deltaMC}
 \delta_n = \sum_{i=1}^n\Lambda_i \sum_{j=i}^n \frac{i^2}{j^2} ,
\end{equation}          
where the $\Lambda_i$ are the independent Gaussian variates of \eqn{pwalk}.  This shows that $\delta_n$ depends only on the steps previous to it.  In the large-$n$ limit, the sum over $j$ simplifies so that 
\begin{equation}
 \label{deltaMClargen}
 \delta_n \simeq \sum_{i=i}^n i\,\Lambda_i \psi_{ni}
          \simeq \sum_{i=i}^n \sqrt{3\Delta s}\, g_i\, (1-i/n)
\end{equation}
where the final expression has used the large $i$ limit for the variance $\avg{\!\Lambda_i^2\!}\to 3\Delta s/i^2$, and the $g_i$ are zero-mean unit-variance Gaussian random numbers.  Thus, we may think of this as a process in which correlations between steps arise because of a smoothing window that has shape $\sqrt{3}(1-S/s)$ when $S\le s$ and is 0 otherwise.  A smoothing window which is a step function leads to Markovian walk heights $\delta$.  For Gaussian fields, this suggests a deeper connection between the $g_i$ and the Fourier modes of the field, which we explore in Section~\ref{kmodes}.  


Before moving on, we note that the model above has $\avg{\!\delta_i\delta_n\!} = (3 - i/n)/2$ for $i\le n$.  One gets the same expression by setting $\kappa=1/2$ in equation~(90) of \cite{mr10a}.  This is not just coincidence.  Since $\kappa=1/2$ is in any case required in their expressions to correctly reproduce the $S\to 0$ limit (in which $f_\mathrm{PS}$ is exact; their value $\kappa\simeq 0.45$ comes from a numerical fit), one might argue that our toy model is in fact the most natural leading order solution of the path integral approach; using walks with uncorrelated steps (Markovian heights) as the leading order approximation led them to significant complications.  As we show below, our rather different approach to this problem is far more accurate and yields much more insight.  

\begin{figure}
 \centering
 \includegraphics[width=0.9\hsize]{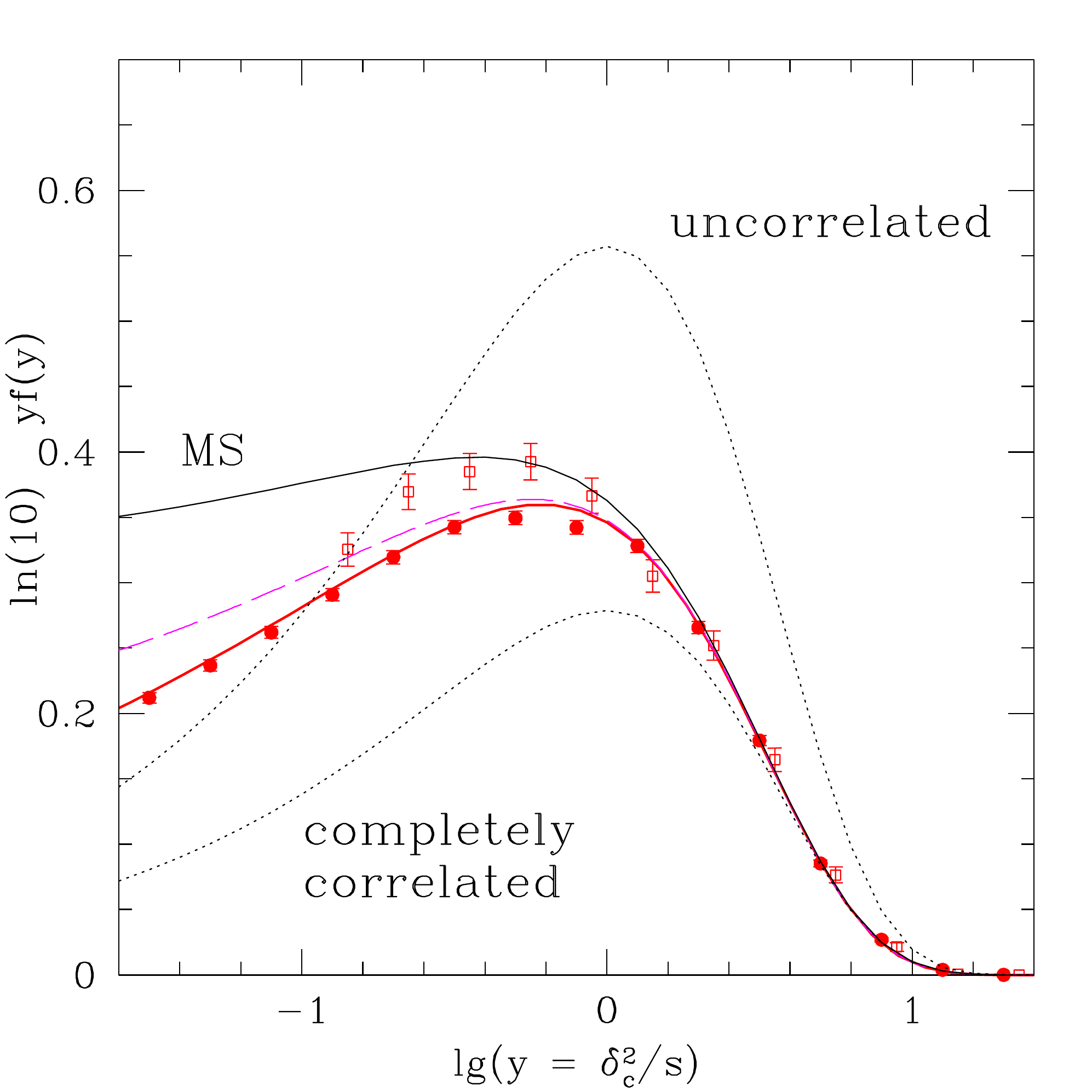}
 \caption{\label{fig:ftoy}
 First crossing distribution $yf(y) = sf(s)$ of a constant barrier of height $\delta_c$, by walks with $\avg{\!v_i v_j\!}=s_i/s_j^2$ for $s_i<s_j$.  Filled circles show the Monte Carlo'd distribution; thick solid curve shows the full back-substitution expression of \citet{ms13b}; dashed curve shows \eqn{ftoy}; and thinner solid curve shows the simpler approximation, $sf_\MS(s)$ of \eqn{fms}.  The two dotted curves show the approximation of \citet{ps74} and twice this value. Open squares show the result of Monte Carloing walks associated with TopHat smoothing of a $\Lambda$CDM power spectrum.}
\end{figure}

\subsection{First crossing distribution}
We noted that our simple model has $\gamma = 1/2$ for all scales.  This value is especially significant for cosmology, because TopHat smoothing of a $\Lambda$CDM power spectrum has $\gamma\sim 1/2$.  Therefore, one might ask if the first crossing distributions are similar.  Figure~\ref{fig:ftoy} compares the first crossing distributions of a barrier of constant height obtained from Monte Carlo realizations of the walks for the process we have just described (filled circles) with several theoretical predictions for the same model and with Monte Carlo realizations of TopHat-smoothed $\Lambda$CDM walks (open squares).  To guide the eye, the two dotted curves show $sf_\mathrm{PS}(s) = (\nu/2)\,{\rm e}^{-\nu^2/2}/\sqrt{2\pi}$ (the original result of \citealt{ps74}) and $2\,sf_\mathrm{PS}$ (the result of \citealt{bcek91} for uncorrelated steps), the thin solid curve shows $f_\MS$ from \eqn{fms} with $\gamma=1/2$, 
and the thick solid curve shows the full back-substitution solution described by \citet{ms13b} (i.e., set $\gamma=1/2$, $\xi = \sqrt{S/s}\,(3 - S/s)/2$ and 
 $\Sigma/(\Gamma\xi) = (1-S/s)/(1 - S/3s)$ in their equations~19 and 20).  
The agreement between the two sets of symbols at large $\delta_c^2/s$ is encouraging.  Nevertheless, the fact that $\gamma$ is mildly scale dependent for $\Lambda$CDM motivates a generalization of the model to accommodate this scale dependence.  We provide this in the next section.  

\begin{figure}
 \centering
 \includegraphics[width=0.9\hsize]{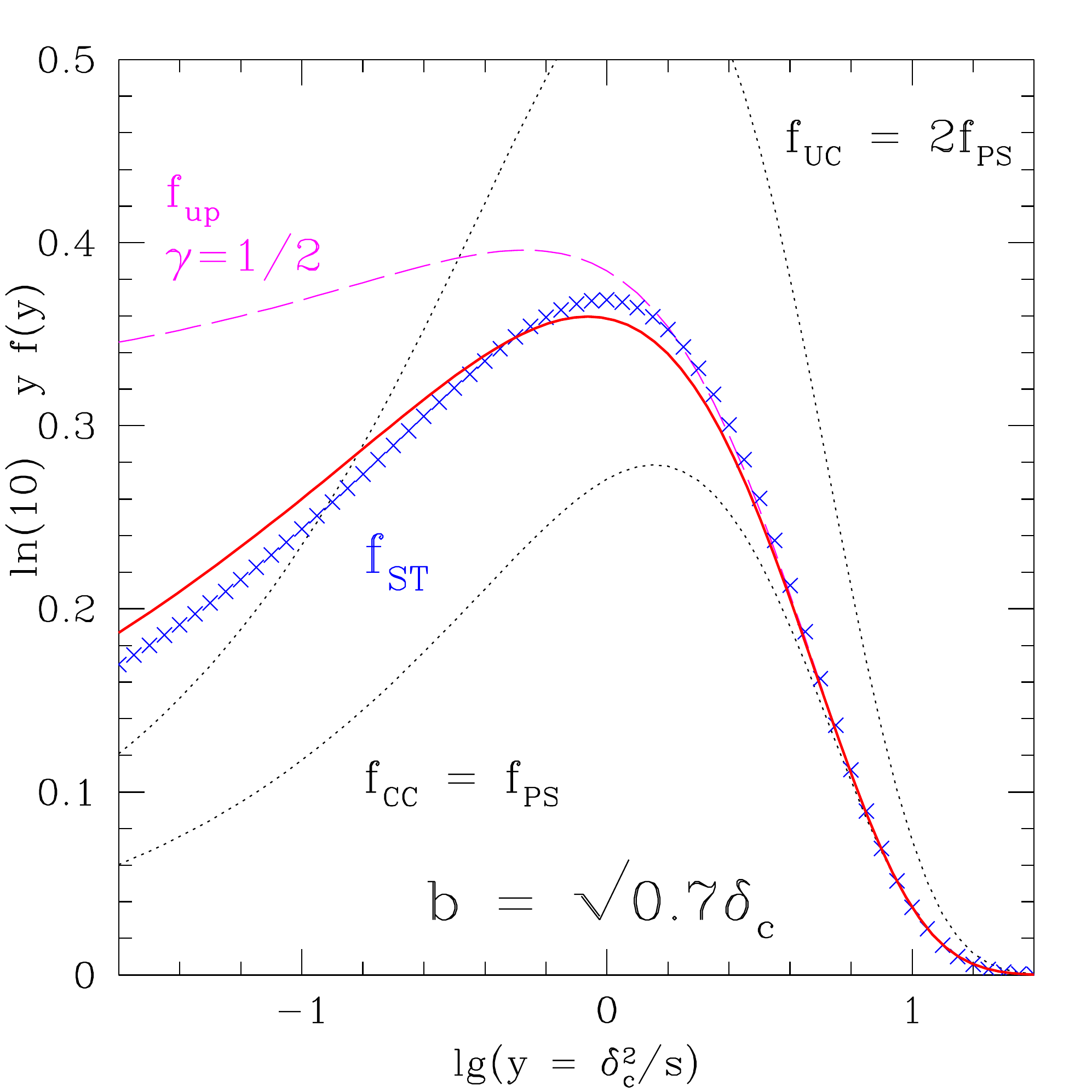}
 \caption{\label{fig:comparest}
 First crossing distribution $yf(y) = sf(s)$ of a constant barrier of height $\sqrt{0.7}\delta_c$, by walks with $\avg{\!v_i v_j\!}=s_i/s_j^2$ for $s_i<s_j$.  Thick solid curve shows the full back-substitution expression of \citet{ms13b}; the previous figure showed that it provides an excellent description of the Monte Carlo'd distribution.  Dashed curve shows the simpler approximation, $sf_\MS(s)$ of \eqn{fms}.  The two dotted curves show the approximation of \citet{ps74} and twice this value. Crosses show the distribution associated with haloes in numerical simulations of structure formation in a $\Lambda$CDM universe.  Our toy model provides a remarkably good description of the crosses.}
\end{figure}

Before doing so, we think it is interesting that our simple toy model actually provides a rather good description of the mass fraction in objects identified in numerical simulations of structure formation in a $\Lambda$CDM universe.  The crosses in Figure~\ref{fig:comparest} show the well-tested fitting formula for this quantity, from \cite{st99}:  
$yf_\mathrm{ST}(y) = 0.322\,[1 + (s/b^2)^{0.3}]\,{\rm e}^{-b^2/2s}/\sqrt{2\pi\,s/b^2}$.  
We argue later that this agreement motivates use of our toy model for generating fast Monte Carlo realizations of halo merger history trees because its walks have correlated steps whose first crossing distribution is a good match to measured halo abundances.


\section{More general models: Markovian velocities}
\label{general}

The analysis of the previous section is specific to the particular form we assumed for the correlation matrix, for which $\avg{\!v_iv_j\!}=s_i/s_j^2$ for $s_i<s_j$.  One could have instead assumed that, for instance, the velocity correlation does not decrease with $s_i$ but remains constant. This happens if $C_{ij}=\ds/(2j+1)$ for $i\leq j$, or $\avg{\!v_iv_j\!}=1/2s_j$ in the continuum limit.  In this case too the same analysis would go through, the only difference being the value of $\psi_{nk}$. Therefore, the question arises as to whether the same can be done in more generality: one can see in fact that the same structure of \eqn{condprob} arises for any process for which $\psi_{nk}$ as defined by \eqn{psitoy} does not depend on $s_i$, which happens for all models for which the dependence of $\avg{\!v_iv_j\!}$ on $s_i$ and $s_j$ factorizes.

We have already noticed that the factorization of the probability of the trajectory in \eqn{pwalk} implies that the steps (and therefore in the continuum limit the velocities) are a Markov process, because the conditional probability of a step only depends on the one before; this is not true for the walk heights (the sums of the steps).  In contrast, for sharp-$k$ smoothing it is the walk heights that are a Markov process, while the steps are uncorrelated (the probability of each step is independent of the previous ones).  We have also argued that the small differences between the toy model and TopHat-smoothed $\Lambda$CDM in Figure \ref{fig:ftoy} are due to the fact that, for $\Lambda$CDM, $\avg{\!v^2\!}$ differs from $1/s$. Therefore, it is interesting to see if one can capture a more general scaling of $\avg{\!v^2\!}$, while preserving the nice features of a Markov process.

\subsection{Langevin equation}
A generic Markov process can be obtained as the solution of a Langevin equation sourced by a white noise $\eta$, such that $\avg{\!\eta(s)\eta(S)\!}=\delta_{\mathrm{D}}(s-S)$.  Let us consider a linear Langevin equation of the form
\begin{equation}
  \label{langevin}
  v'+\frac{\phi'}{\phi}v=\frac{\sqrt{\Phi'}}{\phi}\eta ,
\end{equation}
where the reason for the form of the functions in front of $v$ and $\eta$ will become clear shortly.  This general equation is solved by 
\begin{equation}
  v(s) = \frac{1}{\phi(s)}\int_0^s\dd S\, \sqrt{\Phi'(S)}\,\eta(S) ,
\label{langsol}
\end{equation}
for completely generic functions $\phi(s)$ and $\Phi(s)$.
The white-noise properties of $\eta$ mean that for $S\leq s$
\begin{equation}
  \label{vVFff}
  \avg{\!v(s)\,v(S)\!} =
  \frac{1}{\phi(s)\phi(S)} \int_0^S\dd t\,\Phi'(t)
                       = \frac{\Phi(S)}{\phi(s)\phi(S)}\,,
\end{equation}
which indeed has the required factorized form.

To set $\Phi$, recall that we are only interested in models that have \begin{equation}
  \avg{\!v(s)\delta(s)\!} =
  \int_0^s\dd S\,\avg{\!v(s)v(S)\!} = \frac{1}{2}\,.
\end{equation}
This implies that 
\begin{equation}
 \label{Fs}
  \Phi(S) = \frac{\phi(S)\,\phi'(S)}{2} = \frac{[\phi^2(S)]'}{4}\,,  
\end{equation}
making 
\begin{equation}
  \label{vV}
  \avg{\!v(s)\,v(S)\!}
                       = \frac{\phi'(S)}{2\phi(s)}\,.
\end{equation}
This expression for the correlation between velocities at two different scales generalizes the simple model discussed earlier for the off-diagonal elements of the correlation matrix $C_{ij}$, for which we had set $\avg{\!v_iv_j\!} = s_i/s_j^2$ (recall $s_i\le s_j$). Since this expression is factorized, we can already see that \eqn{condprob} remains true in this very general case also. Of course, we have yet to specify $\phi$.  

To do so, we note that setting $S=s$ in \eqn{vV} implies $2\avg{\!v^2(s)\!}=\dd\ln \phi(s)/\dd s$.
Therefore
\begin{equation}
  \phi(s) = \exp\!\left[2\!\int_{s_0}^s \!\dd t\, \avg{\!v^2(t)\!}\right],
\end{equation}
where $s_0$ is a constant whose exact value does not matter:  it always cancels out because \eqn{langevin} really only depends on $\phi'/\phi$ and $\Phi'/\phi^2 = [\phi^2]'/4\phi^2 = (\phi'/\phi)'/2 + (\phi'/\phi)^2$.  This also means that \eqn{langevin} can be expressed entirely in terms of \avg{\!v^2(s)\!}:  
\begin{equation}
  v' + 2\avg{\!v^2\!}\,v =
  \sqrt{4\avg{\!v^2\!}^2 + \avg{\!v^2\!}'}\,\eta\,,
\label{langevinVsq}
\end{equation}
showing how requiring the Markov process to reproduce the desired $\avg{\!v^2\!}$ and to satisfy $\avg{\!\delta v\!}=1/2$ determines both $\phi$ and $\Phi$, and hence completely fixes $\avg{\!v(s)v(S)\!}$ and the evolution of $v$ itself.  The term under the square root sign on the right-hand is easy to remember:  it is the variance of the quantity on the left-hand side.  Hence, the most natural way to write \eqn{langevinVsq} is to normalize the left-hand side by its variance, so that $\eta$ is obviously unit variance as well.  

What does all this imply for the walk heights $\delta(s) = \int_0^s \dd S\,v(S)$? Since $v$ is itself an integral, integrating \eqn{langsol} and reversing the order of integrations yields 
\begin{equation}
  \delta(s) = \int_0^s\!\dd S \sqrt{\Pi(S)}\,\psi(s,S)\,\eta(S)
\label{genMC}
\end{equation}
with 
\begin{equation}
  \psi(s,S) \equiv 
  \frac{\phi(S)}{S}\!\int_S^s\!\frac{\dd t}{\phi(t)}
  =2\bigg[\frac{\avg{\!\delta(s)\delta(S)\!}}{S}-1\bigg]\,,
\label{genpsi}
\end{equation}
where the last equality comes from integrating the two variables of \eqn{vV} in the $[0,S]$ and $[S,s]$ range respectively, and
\begin{equation}
  \Pi(S) \equiv \frac{\Phi'(S)}{\phi^2(S)}S^2
  = \left[4\avg{\!v^2(S)\!}^2\! +\! \avg{\!v^2(S)\!}'\right]S^2\,.
   \label{Wpsi}
\end{equation}
%
Equation \eqref{genMC} generalizes \eqn{deltaMC}, where the correspondence is recovered upon setting $g_i/\sqrt{\ds} \to \eta(S)$;
this suggests that it may be helpful to think of $\psi(s,S)$ as a window that smooths the white noise $\eta$ to yield $\delta$, and of $\Pi(S)$ as the analogue of the power spectrum. One can check that for the simple model described earlier, for which $\avg{\!v^2\!}=1/s$ and $\phi(s)\propto s^2$, one has $\Pi(S)=3$ and $\psi(s,S)=1-S/s$.

Integrating $v$ between $S$ and $s$ yields $\delta(s)-\delta(S)$. Using \eqn{langsol} for $v$ one then gets
\begin{equation}
  \delta(s) -\delta(S) -\psi(s,S)Sv(S)=
  \int_S^s\!\dd t 
  \sqrt{\Pi(t)}\,\psi(s,t)\eta(t)\,,
\label{cond}
\end{equation}
which generalizes \eqn{condlargestep}. Since the integral runs between $S$ and $s$, and $\eta$ is white noise, this quantity is independent of $\delta(t)$ and $v(t)$ for all $t\leq S$. Its probability is then the conditional probability of $\delta\equiv\delta(s)$ given $\Delta\equiv\delta(S)$ and $V\equiv v(S)$:
\begin{equation}
  p(\delta|\Delta,V,\dots) = p(\delta - \Delta -\psi SV)\,,
\label{probd|DV}
\end{equation}
where the dots stand for any other combination of $\delta(t)$ and $v(t)$ at $t\leq S$. This is a Gaussian with mean 
\begin{align}
 \avg{\!\delta|\Delta,V,\dots} &= \Delta + \psi(s,S)\,SV \,.
\label{d|DV}
\end{align}
Moreover, $\avg{\!(\delta -\Delta-\psi SV)V\!}=0$ implies that 
\begin{equation}
  \psi(s,S)= \frac{\avg{\!(\delta-\Delta)V\!}}{S\avg{V^2}}
\label{genpsi2}
\end{equation}
just like in \eqn{psitoy}.  The variance of $p(\delta|\Delta,V,\dots)$, which we call $C_{\delta\delta|\Delta V}$, is that of $\delta-\Delta-\psi S V$, that is
\begin{equation}
  C_{\delta\delta|\Delta V} =
  \avg{\!(\delta-\Delta)^2\!} 
  - \frac{\avg{\!(\delta-\Delta)V\!}^2}{\avg{V^2}}\,,
 \label{Cdd|DV}
\end{equation}
showing that for Markov Velocities the conditional distribution of $\delta$ at $s$ given the whole trajectory prior to $S$ is simply the conditional distribution of the step $\delta-\Delta$ given $V$.
Therefore, \eqn{condprob} remains true, with $\psi$ now given by \eqn{genpsi2}.

Finally, we note that when $\psi\to 0$ then the walk height $\delta$ is Markovian, since $\delta-\Delta$ becomes an independent variable. (The most generic case, in which it is $\delta-[\avg{\!\delta\Delta\!}/S]\Delta$ that is independent, cannot be recovered from a Langevin equation for $v$ alone.) If instead $\psi$ is finite, equations \eqref{genpsi} and \eqref{genpsi2} imply that
\begin{equation}
  \avg{\!\delta(s)\delta(S)\!} =
  S\left[1+2\gamma_S^2\avg{\!(\delta-\Delta) V\!}\right],
\label{mvSx}
\end{equation}
with $\gamma_S^2 = (4S\avg{\!V^2\!})^{-1}$ as defined in \eqn{gamma}, which recovers the sharp-$k$ correlation function $\avg{\!\delta(s)\delta(S)\!}=S$ when the second term vanishes. This happens not only when $\avg{\!\delta V\!}=\avg{\!\Delta V\!}=1/2$, but also when $\gamma_S=0$, that is when $\avg{\!V^2\!}$ diverges. However, we stress that while if $\delta$ is Markovian then necessarily $\gamma=0$, the converse is not always true: $\gamma=0$ only implies that $p(\delta|\Delta,V)=p(\delta|\Delta)$, and not that  $p(\delta|\Delta,V,\dots)=p(\delta|\Delta)$. Nevertheless, in order to compute $f(s)$ with the approximations used in this paper (quite good, as we have seen), processes with the same $\gamma$ are barely distinguishable (especially for small $\gamma$, see Figure \ref{fig:manyg}), and assuming that $\delta$ is Markovian is still satisfactory.

The expressions above show that $p(\delta|\Delta,V,\dots)$ differs from walks that are Markovian in $\delta$ because of the term $\psi SV$.  However, because $\avg{\!V|\Delta\!} = \Delta\,\avg{\!V\Delta\!}/\avg{\!\Delta^2\!} = \Delta/2S$, we could also have written $\avg{\!\delta|\Delta,V,\dots}$ as $\avg{\!\delta|\Delta\!} + \psi(s,S)\,S\,(V-\avg{V|\Delta})$, which 
instead emphasizes the connection to walks with completely correlated steps.  In this case, the dependence of $p(\delta|\Delta,V,\dots)$ on $\delta-\avg{\!\delta|\Delta\!}$ is the piece identified by \cite{bcek91} (their equation~5.8) and highlighted by \cite{pls12} (their equation~22); our expression shows that there is an additional piece which depends on how far $V$ is from its mean value.

Similarly, one can write \eqn{langsol} as
\begin{equation}
  v(s)-\frac{\phi(S)}{\phi(s)}v(S) = \int_S^s \!\dd t 
 \frac{\sqrt{\Phi'(t)}}{\phi(s)} \eta(t)\,.
\label{vcond}
\end{equation}
The right hand side is independent of both $\Delta$ and $V$, indicating that also
\begin{equation}
  p(v|\Delta,V,\dots)= p\bigg(v - \frac{\phi(S)}{\phi(s)}V\bigg)
\end{equation}
is Gaussian with mean
\begin{equation}
 \avg{\!v|\Delta,V,\dots}
      = \frac{\phi(S)}{\phi(s)}\,V 
      = \frac{\avg{\!vV\!}}{\avg{\!V^2\!}}\,V
      =\avg{\!v|V\!}\,,
\label{v|DV}
\end{equation}
as one could also have seen by differentiating \eqn{d|DV} with respect to $s$, and variance
\begin{equation}
  C_{vv|\Delta V} 
  = \avg{\!v^2\!}- \frac{\avg{\!vV \!}^2}{\avg{\!V^2\!}}\,.
 \label{Cvv|DV}
\end{equation}
That $p(v|\Delta,V,\dots)$ does not depend on $\Delta$ indicates that 
the correlation between $v$ and $\Delta$ arises entirely because of the individual correlations between $v$ and $V$, and $\Delta$ and $V$.  That is, in these models, 
\begin{equation}
 \avg{\!v\Delta\!} = \avg{\!vV \!}\avg{\!V\Delta \!}/\avg{\!V^2\!}\,,
\end{equation}
as it can be seen directly by taking the correlation of \eqn{vcond} with $\Delta$.

Later on we will also need the conditional probability $p(\delta|v,\Delta,V,\dots)$, which is the distribution of a linear combination of $\delta-\Delta-\psi SV$ and $v-\avg{\!v|V\!}$ that is also uncorrelated with the latter. This combination is
\begin{equation}
  \delta-\Delta-\psi SV - \frac{C_{\delta v|\Delta V}}{C_{vv|\Delta V}}
  \bigg(v-\frac{\avg{\!vV\!}}{\avg{\!V^2\!}}\,V \bigg),
\end{equation}
where $C_{\delta v|\Delta V}$ is the covariance of the two variables, and is a Gaussian variate with zero mean and variance 
\begin{equation}
  C_{\delta\delta|v\Delta V} = 
  C_{\delta \delta|\Delta V} -\frac{C_{\delta v|\Delta V}^2}{C_{vv|\Delta V}}\,.
\end{equation}
Similarly, $p(v|\delta,\Delta,V,\dots)$ is the distribution of
\begin{equation}
  v-\frac{\avg{\!vV\!}}{\avg{\!V^2\!}}\,V 
  - \frac{C_{\delta v|\Delta V}}{C_{\delta\delta|\Delta V}}(\delta-\Delta-\psi SV)\,,
\label{v|dDV}
\end{equation}
having variance 
\begin{equation}
  C_{vv|\delta\Delta V} = 
  C_{vv|\Delta V} -\frac{C_{\delta v|\Delta V}^2}{C_{\delta \delta|\Delta V}}\,.
\end{equation}
Finally, since $v-\avg{\!v|V\!}$ is uncorrelated with both $\Delta$ and $V$, $C_{\delta v|\Delta V}$ is just its covariance with $\delta$, that is
\begin{equation}
 C_{\delta v|\Delta V}
 = \frac{1}{2} - \frac{\avg{\!vV\!}}{\avg{\!V^2\!}}\avg{\!\delta V\!}
 = \frac{1}{2} -\avg{\!vV \!}S(2\gamma_S^2+\psi)\,.
 \label{Cdv|DV}
\end{equation}


\subsection{Scale-invariant models}
To see what all this implies, recall that the simplest Markovian velocities model had $\avg{\!v^2\!} = 1/s$.  We will call this a scale-invariant model because it makes $\gamma = (4s\avg{\!v^2\!})^{-1/2}$ independent of $s$.  The analysis above shows that to produce models with other (constant) values of $\gamma$ one simply sets $\avg{\!v^2\!} = (4\gamma^2s)^{-1}$ in \eqn{langevinVsq}.  This means that $\phi = (s/s_0)^{1/2\gamma^2}$, so 
\begin{equation}
 \label{vVpowerlaw}
 \avg{\!v(s)v(S)\!} = \frac{(S/s)^{1/2\gamma^2-1}}{4\gamma^2s}
\end{equation}
and, for $\gamma^2\neq 1/2$,
\begin{equation}
 \label{psiPowerlaw}
 \psi(s,S) = \frac{1 - (S/s)^{1/2\gamma^2-1}}{1/2\gamma^2 - 1}\,,
\end{equation}
while for $\gamma^2=1/2$ one finds $\psi(s,S)=\ln(s/S)$.
In addition, 
$\Pi(S) = (1-\gamma^2)/4\gamma^4 = (4\gamma^2\Gamma^2)^{-1}$ is also still independent of $S$, so that
\begin{equation}
 \label{deltaPowerlaw}
 \delta(s) = \int_0^s \dd S\,\eta(S)\, \frac{\psi(s,S)}{2\gamma\,\Gamma}\,,
\end{equation}
which generalizes \eqn{deltaMC}.
Note that $\avg{\!v(s)v(S)\!}$ and $\psi(s,S)$ diverge when $S/s\to 0$ if $\gamma^2> 1/2$. However, by construction $\gamma^2\leq1$, and $\avg{\!\delta(s)\delta(S)\!}\to0$ in this limit.

\begin{figure}
 \centering
 \includegraphics[width=0.9\hsize]{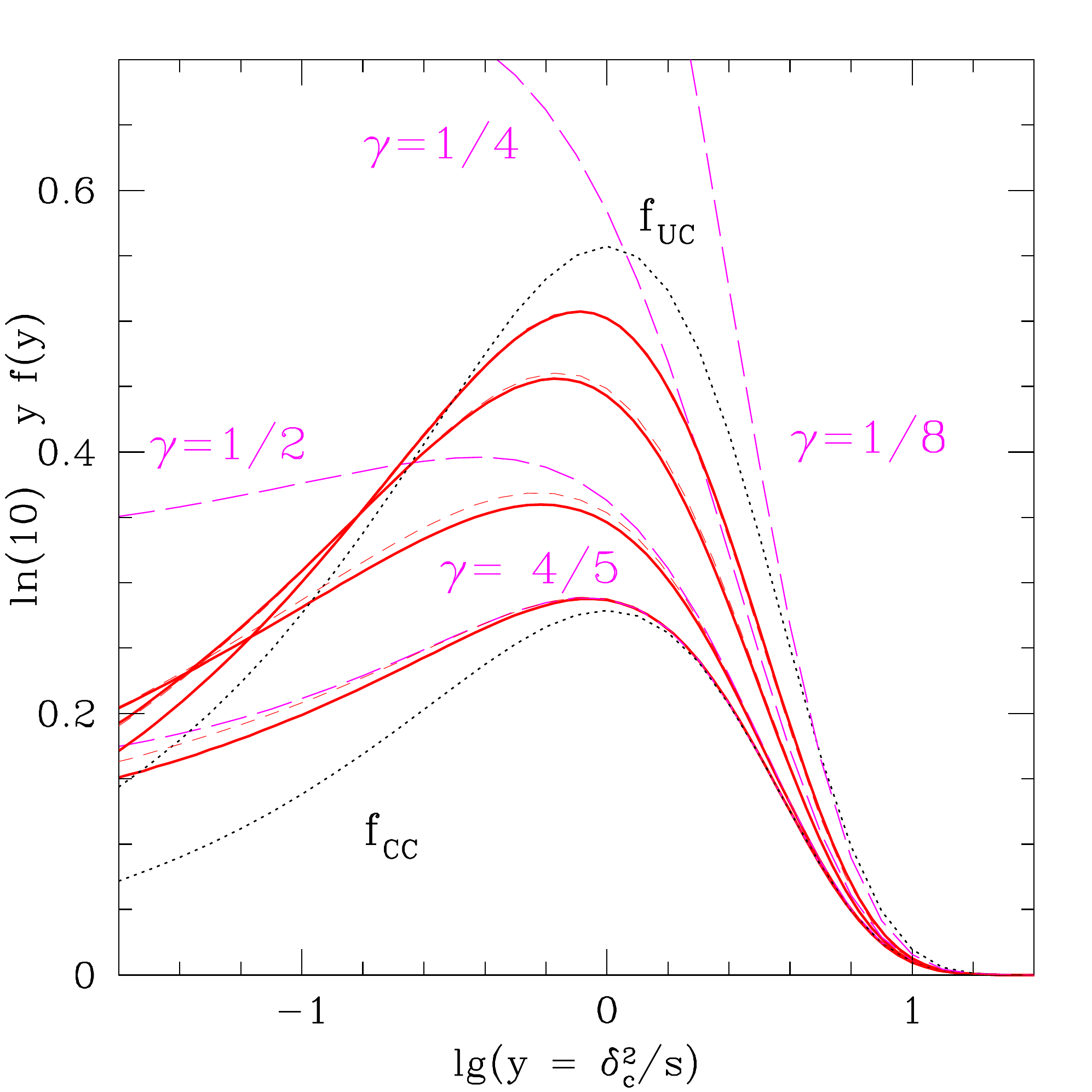}
 \caption{\label{fig:manyg}
 First crossing distribution $yf(y) = sf(s)$ of a constant barrier of height $\delta_c$, by scale-invariant Markov velocity walks with the quoted values of $\gamma$.  The two dotted curves show the approximation of \citet{ps74} and twice this value (the distributions for walks with completely correlated and uncorrelated steps).  Long-dashed curves show $sf_\MS(s)$ of \eqn{fms} and thick solid curves show $sf_\mathrm{BS}(s)$ of \eqn{fbs}.  The latter provide an excellent description of our Monte Carlos in all cases, correctly returning $f_\PS$ and $2f_\PS$ as the limiting cases when $\gamma\to 1$ and 0, respectively, whereas $f_\MS$ becomes an increasingly bad approximation as $\gamma$ decreases.  Short dashed curves show the back-substitution expression for Gaussian smoothed walks that have the same $\gamma$; although they do not have Markov Velocities, their first crossing distributions are very well-approximated by those for Markov velocity walks.}
\end{figure}

This model provides a family of first crossing distributions, indexed by $\gamma$, which interpolate smoothly between the original formula of \cite{ps74} (when $\gamma\to 1$) and twice this formula (when $\gamma\to 0$).  \cite{pls12} identified these as being the limiting cases of walks with completely correlated and uncorrelated steps.  Figure~\ref{fig:manyg} shows this for a few representative choices of $\gamma$.  Whereas $f_\MS$ becomes an increasingly bad approximation to $f(s)$ as $\gamma$ decreases (dashed curves), $f_\mathrm{BS}$ remains well behaved.  Indeed, inserting equations~(\ref{vVpowerlaw}) and~(\ref{psiPowerlaw}) in the back-substitution algorithm of \cite{ms13b} yields excellent agreement with the Monte Carlo'd $f(s)$ (so we have not shown the Monte Carlos).  

We now turn to the question of how well such Markovian Velocity walks approximate walks with Gaussian or TopHat smoothing filters. One would in fact very much like to use Markovian velocity models to mimic generic walks, and, in particular,  to provide a realistic approximation to TopHat smoothed $\Lambda$CDM walks.

A very important case is when $\gamma^2 = 1/6$, since this is also the value for TopHat smoothing of $P(k)\propto k^{-2}$.  For this particular power spectrum and filter, the full correlation structure of the TopHat smoothed walk heights is given by 
\begin{equation}
 \avg{\!\delta(s)\delta(S)\!}
   = S\, [5 - (S/s)^2]/4 \qquad {\rm for}\quad {S\le s}.
 \label{dDn-2}
\end{equation}
This agrees {\em exactly} with the Markov Velocities model:  set $\gamma^2=1/6$ in \eqn{psiPowerlaw} and insert in~(\ref{mvSx}).  
We therefore conclude that TopHat smoothing of $P(k)\propto k^{-2}$ is characterized by truly Markovian walk velocities.  This is one of our key results.  
This case is however special:  as we discuss in Section \ref{kmodes} below, TopHat smoothing of a different $P(k)$ does not lead to Markovian velocities.  

Furthermore, Gaussian smoothing of $P(k)\propto k^n$ has 
 $\avg{\!\delta(s)\delta(S)\!} = S\,[2/(1 + (S/s)^{1/\Gamma^2})]^{\Gamma^2}$, 
where $\Gamma^2 = (n+3)/2$.  Since this is not of the form of \eqn{mvSx} for any $n$, Gaussian smoothed walks never have Markovian velocities.  
For instance, Gaussian smoothing of $P(k) \propto k^{-1}$ yields $\gamma^2=1/2$ and $\avg{\!\delta\Delta\!} =2S/(1 + S/s)$, while the corresponding Markovian velocity model has $\psi = \ln(s/S)$ and thus $\avg{\!\delta\Delta\!} = S[1 - \ln(S/s)/2]$. The two differ by more than 10 per cent for $S/s<0.2$.


Therefore, in order to reproduce a generic correlation structure with Markovian velocities, there are two possible choices:  to match the velocity correlation parameter $\gamma$, or to match the amplitude of the spatial correlations -- the $S/s\to 0$ limit of $\avg{\!\delta(s)\delta(S)\!}/S$ (the $S/s\to 1$ limit is, of course, always equal to unity).  For Gaussian smoothing, the latter condition means $2^{\Gamma_{\rm G}^2} = (1-\Gamma_{\rm MV}^2)^{-1}$:  the two $\Gamma$s are not the same.  
We have found that matching $\gamma$ produces slightly better agreement, most likely  because the integral in \eqn{fbs} (which we solve numerically) receives little contribution at small $S$: so we have only shown these cases -- using short dashed lines -- in Figure~\ref{fig:manyg}.
The agreement is rather impressive, suggesting that Markov Velocities are a rather good approximation to the full story.  



We have thus shown that the first crossing distributions of scale-invariant Markov Velocity models are in good agreement with those of Gaussian or TopHat-smoothed walks having the same $\gamma$, over a wide range of scale-free $P(k)$.  This raises the question of how well a Markovian velocity model can mimic TopHat-smoothed $\Lambda$CDM walks, for which $P(k)$ is not a simple power law.  We address this in the next subsection.

\subsection{Scale-dependent models}
Although the scale-invariant models, being completely analytic, yield considerable insight, they are rather restrictive.  Indeed, because \eqn{langevinVsq} itself does not require $\avg{\!v^2\!}\propto s^{-1}$, it allows us to model cases in which $\gamma$ is a function of scale $s$.  Thus, for example, we can choose the scale dependence of $\avg{\!v^2(s)\!}$ to be exactly the same as that of TopHat-smoothed $\Lambda$CDM walks.  As before, this will not guarantee that the full correlation matrix $\avg{\!v_iv_j\!}$ will also be matched exactly, but, following the discussion above, we expect the resulting walks to provide a good approximation to the $\Lambda$CDM ones.  

In particular, such walks have 
\begin{equation}
 \gamma(s) \approx a + b\ln(s/\delta_c^2)
 \label{gsLCDM}
\end{equation}
with $a=0.45$ and $b=-0.03$.  Recalling that $\avg{\!v^2(s)\!} \equiv (4s\gamma^2)^{-1}$ means 
\begin{equation}
 \Pi(S) = \frac{1 - \gamma^2(S) - 2b\gamma(S)}{4\,\gamma^4(S)}
\end{equation} 
and 
\begin{equation}
 \psi(s,S) = \int_1^{s/S}\dd \tau\,
    \exp\left[-\frac{\ln\tau/\gamma(S)}{2(\gamma(S)+b\ln\tau)} \right].
\end{equation}
If we simply ignore the factor of $b\ln\tau$ in the denominator of the term in the exponential, then 
\begin{equation}
 \label{psiLCDM}
 \psi(s,S) 
  \approx \frac{1 - (S/s)^{1/2\gamma^2(S) - 1}}{1/2\gamma^2(S) - 1},
\end{equation}
which is \eqn{psiPowerlaw} but with $\gamma\to\gamma(S)$.  If we also drop the $2b\gamma(S)$ term from $\Pi$ (recall $b\ll 1$), then even the `power spectrum' is simply given by replacing $\gamma \to\gamma(S)$.  Therefore, \eqn{deltaPowerlaw} with $\gamma\to\gamma(S)$ can be used to generate Monte Carlo realizations of walks.  

Note that at $s=1$, $\gamma = 0.48$ and $\Pi = 3.7$, which are not far from our original simplest model which had $\gamma=1/2$ and $\Pi = 3$, and which Figure~\ref{fig:ftoy} showed worked reasonably well at $s\le 1$.  Since our $\gamma$ decreases as $s$ increases, we expect the associated first crossing distributions to become more like those for smaller $\gamma$ (i.e., more negative $n$).  

\subsection{MVMC:  Markov Velocity Monte Carlo}
In practice, we do not generate walks using the approximate expression for $\psi$ given in \eqn{psiLCDM}.  Rather, we generate walks with the full structure (i.e. no dropping of $b\ll 1$ terms) by replacing \eqn{deltaMC} with 
\begin{equation}
 \delta_n = \sum_{j=1}^n \sum_{i=1}^j \sqrt{\Pi_i\,\Delta s}\, g_i\,
             \frac{i}{j^2} \,
        \frac{(S_i/\delta_c^2)^{1/2a\gamma_i - 2}}{(S_j/\delta_c^2)^{1/2a\gamma_j - 2}}
 \label{deltaLCDM}
\end{equation}
where $S_k = k\,\Delta s$.


\begin{figure}
 \centering
 \includegraphics[width=0.9\hsize]{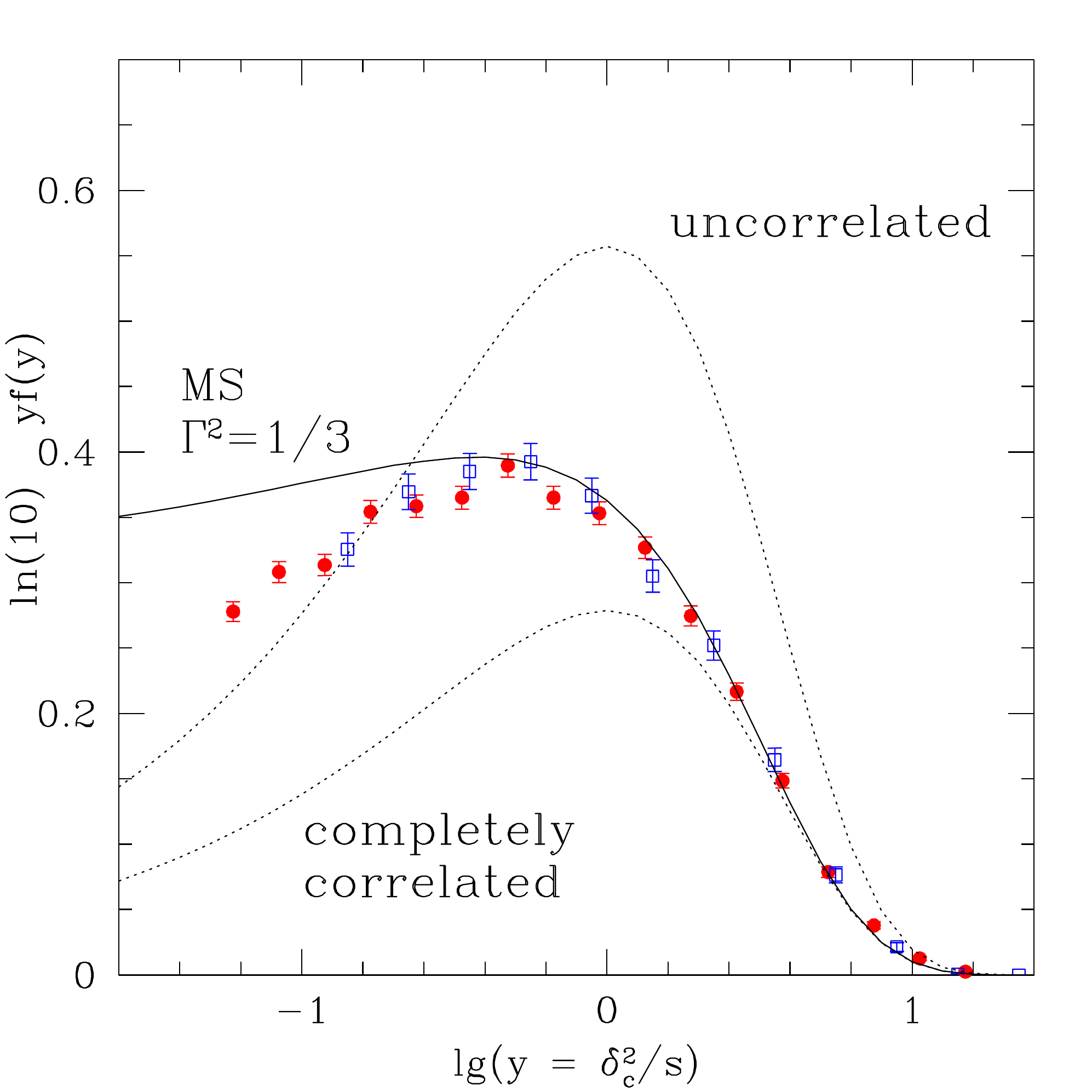}
 \caption{\label{fig:markov2lcdm}
 Comparison of the first crossing distribution of a constant barrier of height $\delta_c$, by walks associated with the exact TopHat smoothing of a $\Lambda$CDM power spectrum (open squares, same as previous Figure), and by walks that have the same $\avg{\!v^2(s)\!}$ but Markovian velocities (filled dots).  For comparison, the thin solid curve shows the simpler approximation of \citet{ms12} with $\Gamma^2=1/3$ (same as previous Figure) and the two dotted curves show the approximation of \citet{ps74} and twice this value. }
\end{figure}

Figure~\ref{fig:markov2lcdm} shows a comparison of the first crossing distribution of a constant barrier of height $\delta_c$, by walks associated with TopHat smoothing of a $\Lambda$CDM power spectrum (open squares, same as previous figure), and for walks which have the same $\avg{\!v^2(s)\!}$ but Markovian velocities.  For comparison, the thin solid curve shows the simpler approximation of \citet{ms12} with $\Gamma^2=1/3$ (same as previous figure) and the two dotted curves show the approximation of \citet{ps74} and twice this value.  Clearly, our Markov velocities model provides an excellent approximation to the actual distribution.  

Fundamentally, this is because the velocity correlations are largest on the diagonal of $C_{ij}$, both for the Markov Velocities model and the actual TopHat smoothed walks.  However, the toy model has a sharp edge while the real thing is smooth.  By ensuring the Markov model has the same $\avg{\!v^2(s)\!}$, we are forcing the terms along the diagonal $C_{ii}$ to be the same, but because $\avg{\!\delta v\!} = 1/2$, the Markov model $C_{ij}$ is less than the real one for $S\sim s$, but is larger for $S\ll s$, so the two errors approximately compensate.  This, -- i.e., the need to set $\avg{\!v^2(s)\!}$ to that of the real walks -- is why it was necessary to go beyond the simplest toy model we developed initially.  





\subsection{Markov velocities and Fourier space filters}
\label{kmodes}

We remarked earlier that one could think of the Markov Velocities model as one in which there is a smoothing kernel $\psi(s,S)$ applied to a power spectrum $\Pi(S)$.  However, \eqn{sR} shows that one can always think of $\delta(s)$ as a weighted sum over Fourier modes, with the variance on scale $k$ given by the power spectrum.  What does our Markov Velocities model imply for the window that smooths $P(k)$?  

Let us first consider filters that are compact in Fourier space. For instance, if we set 
\begin{equation}
 \label{Wtruncated}
 W_\alpha(kR) \equiv [ 1 - (kR)^\alpha]\, \vartheta(1-kR)\,,
\end{equation}
then for power-law power spectra, $P(k)\propto k^n$, the integrals which define $S(R)$ and $\avg{\!\delta(s)\delta(S)\!}$ can be done analytically, from which $\avg{\!v(s)v(S)\!}$ can also be derived.  Comparison with the scale-invariant results derived previously shows that applying our truncated filters $W_\alpha$ of \eqn{Wtruncated} to power-law $P(k)$ will result in a Markovian Velocity model having 
\begin{equation}
 1/\gamma^2 = 2 + 2\alpha/(3+n).
 \label{an2gamma}
\end{equation}
Clearly, for more general $P(k)$, such as for the $\Lambda$CDM family, where the mapping between $R$ and $S$ is not a power law but is still monotonic, the shape of $\psi$ depends on both $s$ and $S$ rather than $S/s$, as well as on $\alpha$.

However, regardless of the shape of $P(k)$, differentiating \eqn{Wtruncated} twice gives
\begin{equation}
  \frac{\dd^2  W_\alpha}{\dd R^2} 
  + \frac{1-\alpha}{R}\, \frac{\dd W_\alpha}{\dd R} = 
  \frac{\alpha}{R^2} \delta_{\mathrm{D}} (1-kR)\,,
\end{equation}
where now the right hand side only selects those Fourier modes with $k=1/R$. It follows that the field filtered with $W_\alpha(kR)$ satisfies 
\begin{equation}
    \frac{\dd^2 \delta_\alpha}{\dd R^2}
  + \frac{1-\alpha}{R}\, \frac{\dd \delta_\alpha}{\dd R} 
  = \eta_\alpha,
\label{lang_alpha}
\end{equation}
where $\eta_\alpha$ is white noise.  This is a Langevin equation for $\dd \delta_\alpha/\dd R$, which gives another Langevin equation for $v$ when changing variables from $R$ to $s$. Therefore, whatever the shape of $P(k)$, one can build a family of Markovian velocity models indexed by the value of $\alpha$. The usual Markovian walk heights are recovered in the limit $\alpha\to\infty$, when the second derivative becomes irrelevant.

That Markovian velocities may arise from truncated Fourier filters makes intuitive sense, but the underlying reason is that the derivative is discontinuous.  For example, filters of the form $(1 - |kR|)^\alpha\, \vartheta(1-kR)$, while truncated, do not yield Markovian velocities because for them, the discontinuity appears in higher derivatives (e.g., $\alpha=2$ yields Markovian accelerations, and so on). On the other hand, filters of the form $|kR|^\alpha\,\vartheta(1-kR)$ are already discontinuous and yield directly Markovian heights with $\avg{\!\delta(r)\delta(R)\!} = \avg{\!\delta^2(R)\!} (r/R)^\alpha$ (the usual sharp-$k$ smoothing is the $\alpha\to 0$ limit of this more general Markov process). In general, if the $n$-th derivative of the filter is discontinuous, the $n$-th derivative of $\delta$ will be Markovian.

Even though  the truncated filters in \eqn{Wtruncated} can reproduce all possible (constant) values of $\gamma$, they are not the only ones which give rise to walks with Markov Velocities. For instance, smoothing $P(k)\propto k^{-2}$ with the TopHat filter 
\begin{equation}
  W_{\mathrm{TH}}(kR)=(3/kR)\,j_1(kR)
\end{equation}
yields the same correlation structure (equation~\ref{dDn-2} and $\gamma^2 = 1/6$) as the model obtained using \eqn{Wtruncated} with $\alpha=2$. 
It is then interesting to see when a TopHat filter, which has great physical interest but is clearly not truncated nor discontinuous in $k$, gives a Markov process. 

Spherical Bessel functions satisfy the recurrence relation 
$x^m[-(1/x)(\dd/\dd x)]^m[j_1(x)/x]=j_{m+1}(x)/x$, from which it can be seen that a TopHat-filtered $\delta_{\mathrm{TH}}$ obeys the differential equations
\begin{gather}
  \frac{\dd \delta_{\mathrm{TH}}}{\dd R} 
  =  \eta_1\qquad \mathrm{and} \qquad
  \frac{\dd^2  \delta_{\mathrm{TH}}}{\dd R^2} 
  - \frac{1}{R}\, \frac{\dd \delta_{\mathrm{TH}}}{\dd R} = 
  \eta_2\,.
\end{gather}
Upon noting that $\delta_{\mathrm{TH}}\equiv\eta_0$, the statistics of the field and its derivatives are described by the correlation functions 
\begin{equation}
  \avg{\!\eta_{a\!}(r)\eta_b(R)\!} 
  = \!\int\! \dd k \,\frac{9(-k)^{a+b} P(k)}{2\pi^2rR}  j_{a+1}(kR)j_{b+1}(kr).
\end{equation}
When $a=b=0$, this returns $\avg{\!\delta_{\mathrm{TH}}(r)\delta_{\mathrm{TH}}(R)\!}$. This scheme can be extended to any higher derivative, yielding differential equations for $\dd\delta_{\mathrm{TH}}/\dd R$ of arbitrary order.

For a power-law $P(k)\propto k^n$, one sees that $\eta_1$ behaves like white noise when $n=0$ (as the integral above reduces to the closure equation for spherical Bessel functions), making $\delta_{\mathrm{TH}}$ a Markov process. For $n=-2$, it is $\eta_2$ that becomes white noise, while the correlation function of $\eta_1$ has a factorized form: therefore, it is $\dd\delta_{\mathrm{TH}}/\dd R$, the velocity, which is a Markov process.  And indeed, in this case, the differential equation for $\eta_2$ is the same as $\alpha=2$ in \eqn{lang_alpha}. (For $n=-4$ TopHat smoothing has Markovian accelerations, and so on.)  For all other values of $n$, strictly speaking, neither process is Markovian. However, one can check that for $-1\leq n <0$ the variance $\avg{\!\eta_1^2(R)\!}$ diverges, making $\gamma=0$.  In this case, for single-scale approximations such as the upcross and back-substitution approximations described in this paper, $\delta_{\mathrm{TH}}$ effectively behaves like a Markov process.  In contrast, for $-3\leq n <1$ it is $\avg{\!\eta_2^2(R)\!}$ that diverges:  in these cases, the single-scale approximations will be well described by Markov Velocity models.

\section{Applications to galaxy formation and cosmology}
\label{sec:cond}

As we have seen, Markov Velocities arise quite naturally when smoothing with a TopHat filter a field having power spectrum close to that of  $\Lambda$CDM. This makes these models very useful for studying the evolution and clustering of structures.  In this section, we explore several applications of Markov Velocity models, relying on their key property that the conditional probability of $\delta$ and $v$ at one scale may only depend on $\Delta$ and $V$ at one larger scale.  That is to say, in Markov Velocity models, at fixed halo mass and slope of the initial density profile (measured on the mass scale of the protohalo), there should be no correlation between formation history and the larger scale environment.  
We also show that the converse is also true:  at fixed mass and slope, the density profile on large scales (i.e., the environment) is statistically the same regardless of the shape of the inner profile (i.e., the formation history).

\subsection{Correlations with larger scales}\label{lssbias}
Let $f(s|\Delta,S)$ denote the fraction of walks which first cross $b(s)$ on scale $s$, subject to the constraint that the walk had height $\Delta$ on scale $S<s$.  Then $f(s|\Delta,S)\approx f_\MS(s|\Delta,S)$ where
\begin{equation}
 f_\MS(s|\Delta,S) = p(b|\Delta)\,\int \dd v\,(v-b')\,p(v|b,\Delta)
\end{equation}
\citep{mps12}.  The similarity to \eqn{fms} means that $f_\MS(s|\Delta,S)$ equals $p(b|\Delta)$ times a correction factor which depends on the mean and variance of $p(v|b,\Delta)$.

The conditional distribution $p(v|b,\Delta)$ is a Gaussian with mean 
\begin{equation}
 \avg{\!v|b,\Delta\!} = \avg{\!v|\Delta\!} + 
  \frac{C_{v\delta|\Delta}}{C_{\delta\delta|\Delta}}\,(b - \avg{\!\delta|\Delta\!})
 %
\end{equation}
and variance 
\begin{equation}
 C_{vv|\delta\Delta} = C_{vv|\Delta} - \frac{C_{v\delta|\Delta}^2}{C_{\delta\delta|\Delta}}\,,
\end{equation}
where, for our Markov Velocity models,
\begin{align}
 \avg{\!v|\Delta\!} &=\avg{\!V|\Delta\!}\,\avg{\!vV\!}/\avg{\!V^2\!}\,,\nonumber\\
  C_{v\delta|\Delta} &= \frac{1}{2}\left(1 - \frac{\avg{\!\delta\Delta\!}}{\avg{\!\Delta^2\!}}\,\frac{\avg{\!vV\!}}{\avg{\!V^2\!}}\right), \nonumber\\
  C_{\delta\delta|\Delta} &= s - \avg{\!\delta\Delta\!}^2/S \,, \notag \\
 C_{vv|\Delta} &= \avg{\!v^2\!}\left(1 - \frac{\avg{\!vV\!}^2}{\avg{\!v^2\!}\avg{\!V^2\!}\,4S\avg{\!V^2\!}}\right).
\end{align}
The amount by which the ratio $f(s|\Delta,S)/f(s)$ differs from unity is often used to estimate the bias that comes from the large scale environment; the $S\ll s$ limit, in which $b\gg \avg{\delta|\Delta}$ yields what is known as the bias on large scales.  As discussed in detail by \cite{mps12}, this bias will be a constant in the $k\to 0$ limit, but the dependence on $v$ will lead to $k$-dependent corrections (the leading order term being $\propto k^2$).  Both the $k\to 0$ constant and the amplitude of the $k$-dependent corrections depend on $s$.   




\subsection{Conditional progenitor distributions}\label{hizprog}
The expression above only requires that the walk have height $\Delta$ on scale $S$.  Instead, let $f(s|S)$ denote the conditional distribution of first reaching $b(s)$ on scale $s$ given that the walk first crossed $B(S)$ on scale $S$.  Following \cite{bcek91}, this is the quantity which most excursion set approaches use to model the distribution of progenitors from an earlier time that become part of a more massive object later \cite[e.g.][]{lc93,st02}.  

For walks with uncorrelated steps, $f(s|S)$ has the same form as the unconditional distribution, upon rescaling $\delta\to b-B$ and $s\to s-S$.  Moreover, it also has the same form as $f(s|\Delta=B,S)$ of the previous section.  For walks with correlated steps, however, the analysis is more complicated, and the two quantities are not the same -- the additional `first crossing' constraint on the larger scale matters.  

The same logic that leads to \eqn{fms} for the unconditional distribution suggests that
\begin{equation}
 f(s|S) \simeq f_\MS(s|S),
\end{equation}
where 
\begin{equation}
 f_\MS(s|S) \equiv \frac{p(B)}{f_\MS(S)} \int_{B'}^\infty \!\!\dd V\,(V-B')
              \,p(V|B)\,f_\MS(s|B,V)
 \label{fupsS}
\end{equation}
and 
\begin{equation}
 f_\MS(s|B,V) = p(b|B,V) \int_{b'}^\infty \dd v\, (v-b')\,p(v|b,B,V)
 \label{fupBV}
\end{equation}
is the upcrossing (rather than first crossing) distribution for walks conditioned to start from $B$ with velocity $V$ \cite[see equation~10 and associated discussion in][]{ms12}.  (Strictly speaking, one might replace $f_\MS(s|B,V)$ with the corresponding $f_{\rm BS}(s|B,V)$, for the same reasons one might replace $f_\MS$ with $f_{\rm BS}$ of \eqn{bsApprox}, but we will not do so below.)  This shows that $f_\MS(s|S)$ is a sum over upcrossing distributions $f_\MS(s|B,V)$ conditioned to have different slopes $V$ when they upcrossed, and weighted by the probability of upcrossing with that slope.  

To compute $f_\MS(s|B,V)$, we first note that \eqn{fupBV} has the same structure as \eqn{fms}, with the unconditional distributions there being replaced by conditional ones:   $p(b)\to p(b|B,V)$ and $p(v|b)\to p(v|b,B,V)$. These Gaussian distribution are given by \eqn{probd|DV} and \eqn{v|dDV} respectively.
The similarity to \eqn{fms} means that $f_\MS(s|B,V)$ equals $p(b|B,V)$ times a correction factor that depends on $[b' + \avg{\!v|b,B,V\!}]/C_{vv|\delta\Delta V}^{1/2}$.

Although \eqn{v|dDV} is written in a form which highlights the connection to walks with correlated steps, because $\avg{\!v|\Delta,V\!} = \avg{\!v|V\!}$ does not depend on $\Delta$, $\avg{\!v|\delta,\Delta,V\!}$ is the sum of two terms, one of which is proportional to $V$ and the other to $\delta-\Delta$.  That $\delta$ and $\Delta$ do not appear separately, but only as $\delta-\Delta$ shows the close connection of our Markov Velocities process to one which is Markovian in $\delta$; the additional dependence on $V$ indicates that the walks are non-Markovian in $\delta$.  We made a similar remark when discussing \eqn{d|DV}.  As \cite{ms12} note, this dependence on $V$, this non-Markovian behaviour, shows explicitly that our Markovian Velocities model comes with assembly bias effects built-in.  

Before we discuss these effects, to get some intuition about the shape of $f_\MS(s|S)$, it is instructive to set $V$ equal to its mean value $B/2S$.  This makes $\delta-\avg{\!\delta|\Delta,V\!}\to \delta - \avg{\!\delta|\Delta\!}$.  If we now move $f_\MS(s|B,V=B/2S)$ out of the integral over $V$ in \eqn{fupsS}, this makes the numerator there equal $f_\MS(s|B,V=B/2S)\,f_\MS(S)$, so that $f(s|S) = f_\MS(s|B,V=B/2S)$.  Since $f_\MS(s|B,V=B/2S)\propto p(b|B,V=B/2S)$ at $S/s\ll 1$, the result depends on the scaling variable 
$(b - \avg{\!\delta|B\!})/\sqrt{s(1-S/s)^3}$.
In contrast, the simplest well-motivated approximation for the conditional distribution \cite[equation~22 of][]{pls12} is $(b - \avg{\!\delta|B\!})/\sqrt{C_{\delta\delta|\Delta}}$.  Our analysis provides a simple way to understand why that approximation works rather well:  it boils down to fixing $V$ to its correct mean value and ignoring the fact that the constrained scatter around this mean is narrower than the unconstrained value, but this difference is vanishingly small as $S/s\ll 1$.

\subsection{Assembly bias}\label{abias}
Assembly bias effects associated with Markov Velocity models, while present, are particularly simple.  To see why, consider walks which first crossed the barrier on scale $S$.  Let $\Delta_0$ denote the value of the field on large scales $S_0<S$, and $\delta$ and $v$ the values on some smaller scale $s>S$.  Then 
 $p(\delta,v|\Delta,\Delta_0)$ equals $p(\delta,v|\Delta)$ if the walk heights were Markov, but in general, and for our Markov Velocity walks in particular,  
\begin{equation}
  p(\delta,v|\Delta,\Delta_0) \ne p(\delta,v|\Delta).
  \label{pdvD}
\end{equation}  
The dependence of $\delta$ and $v$ on the large scale $\Delta_0$ is a manifestation of Assembly bias.  

Now consider 
 $p(\delta,v|\Delta,V,\Delta_0)$.  We have already argued that, for Markov Velocity models,
\begin{equation}
 p(\delta,v|\Delta,V,\Delta_0) = p(\delta,v|\Delta,V),
 \label{pdvDV}
\end{equation} 
illustrating that, if $\Delta$ {\em and} $V$ are specified, there is no correlation with the large scale $S_0$.  Therefore, although assembly bias is present (equation~\ref{pdvD}) -- the analysis of the previous subsection showed that the mean values of $p(b|B,V)$ and $p(v|b,B,V)$ in $f_\MS(s|B,V)$ are shifted -- these effects are particularly simple.  In particular, if both $B$ and $V$ have been specified, there are {\em no} additional correlations with the larger scale environment.

Now, write the left-hand side of \eqn{pdvDV} as $p(\delta,v,\Delta,V,\Delta_0)/p(\Delta,V,\Delta_0)$, and the right-hand side as $p(\delta,v,\Delta,V)/p(\Delta,V)$.  Then \eqn{pdvDV} implies that 
\begin{equation}
 \frac{p(\Delta_0|\delta,v,\Delta,V)\,p(\delta,v,\Delta,V)}
      {p(\Delta,V,\Delta_0)} 
 = \frac{p(\delta,v,\Delta,V)}{p(\Delta,V)},
\end{equation}
that is
\begin{equation}
 p(\Delta_0|\delta,v,\Delta,V) = \frac{p(\Delta,V,\Delta_0)}{p(\Delta,V)}
  = p(\Delta_0|\Delta,V),
 \label{pD0DV}
\end{equation}
which is explicitly independent of $\delta$ and $v$.  Note that $p(\Delta_0|\Delta,V)$ is Gaussian, with mean 
\begin{equation}
  \avg{\!\Delta_0|\Delta,V\!} = 
  \frac{\avg{\!\Delta_0\Delta\!}}{S}\Delta + 
  \frac{\avg{\!\Delta_0(V-\Delta/2S)\!}}{\avg{\!V^2\!}(1-\gamma_S^2)}
  \!\left(V - \frac{\Delta}{2S}\right)\!.
\end{equation}
For scale-free $P(k)$ the term which multiplies $V-\Delta/2S = -S_0\psi(S_0,S)$, showing explicitly that if $V > \Delta/2S$ then the dependence on $V$ acts to decrease the large scale $\Delta_0$.  

\allowdisplaybreaks

\begin{figure}
 \includegraphics[width=.45\textwidth]{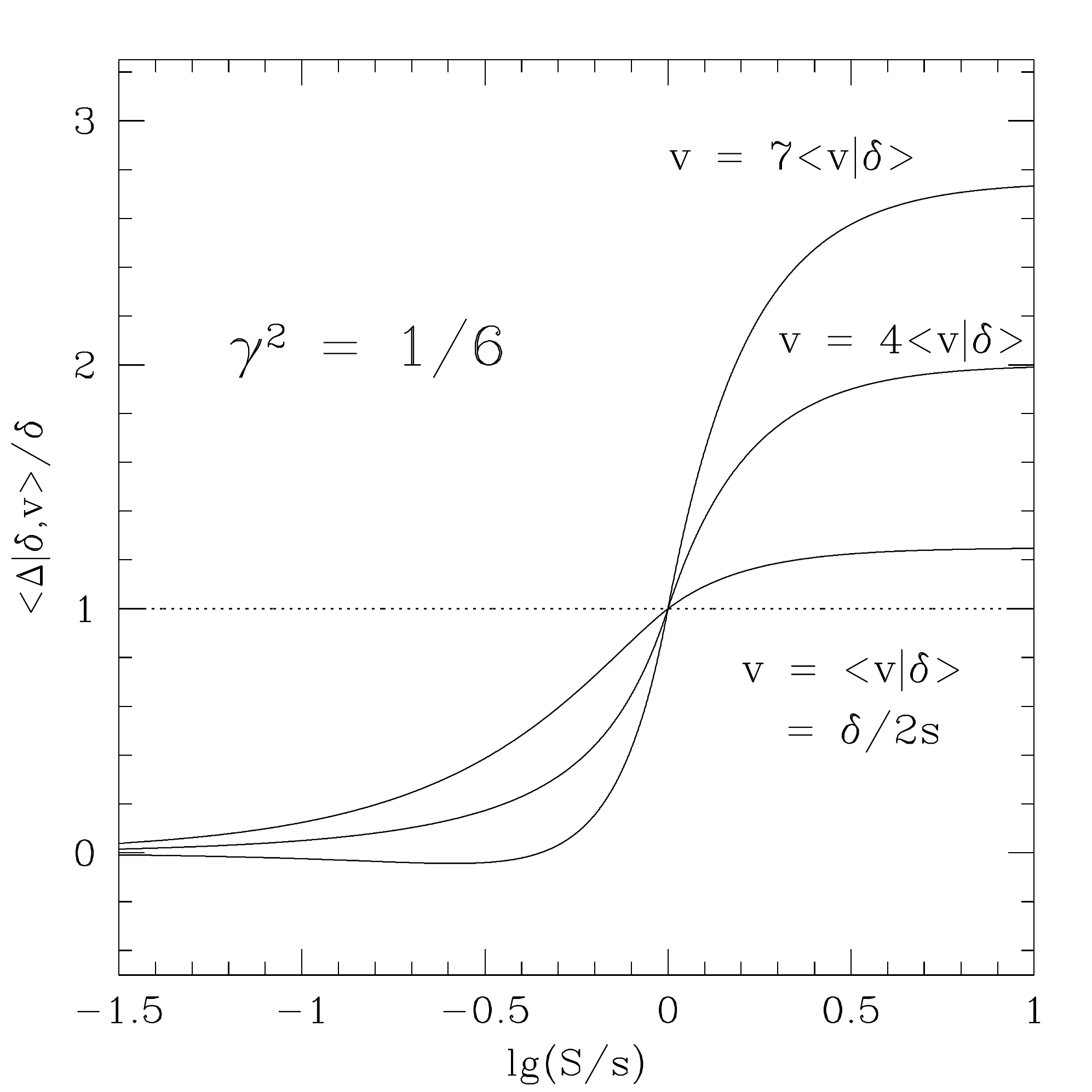}
 \caption{\label{fig:ab}
Assembly bias associated with TopHat smoothing of $P(k)\propto k^{-2}$. Curves show the mean walk height on scale $S$ subject to the constraint that on scale $s$ the walk has height $\delta$ and slope $v$.  The three choices of $v = 1,4,7\times\avg{v|\delta}$ show that steeper walks are associated with less dense large scale environments.}
\end{figure}

To illustrate these effects, Figure~\ref{fig:ab} shows $\avg{\Delta|\delta,v}/\delta$ for a range of choices of $v$, for TopHat smoothing of $P(k)\propto k^{-2}$.  In this case $\gamma^2=1/6$, so
\begin{equation}
 \avg{\!\Delta|\delta,v\!} = \delta\,\frac{S}{4s} \left[5 - \frac{S^2}{s^2} - \left(1 - \frac{S^2}{s^2}\right) \left(\frac{v}{\delta/2s} - 1\right)\right]
\end{equation}
and 
\begin{equation}
 \avg{\!\delta|\Delta,V\!} = \Delta + S\,\frac{1 - (S/s)^2}{2}\,V,
\end{equation}
where our notation is for $S<s$.  The steeper curves in the figure are associated with larger $v$:  clearly, large overdensity at $S/s\ge 1$ is associated with less dense environments (i.e., smaller heights at $S/s\le 1$.  The other quantities which matter for assembly bias related quantities are  
\begin{align}
 \langle v|\Delta,V\rangle &= (S/s)^3\,V, \qquad
 C_{vv|\Delta V} = \frac{1 - (S/s)^5}{2s/3}, \nonumber\\
 C_{\delta\delta|\Delta V} &= s(1 - S/s)^3\left(1 + \frac{9S}{8s} + \frac{3S^2}{8s^2}\right), \nonumber\\
 C_{\delta v|\Delta V} &= \frac{1}{2} - \frac{5S^3}{4s^3} + \frac{3S^5}{4s^5}\,.
\end{align}
To see that these effects differ quantitatively for different Markov Velocity models, note that the simplest toy model ($\gamma=1/2$) has 
\begin{align}
 \avg{\!\Delta|\delta,v\!} &= \delta\frac{S}{2s}\left[3 - \frac{S}{s} - \left(1 - \frac{S}{s}\right)\left(\frac{v}{\delta/2s} - 1\right)\right] , \nonumber\\
 \avg{\!\delta|\Delta,V\!} &= \Delta + S(1 - S/s)\,V\,, \nonumber\\
 \langle v|\Delta,V\rangle &= (S/s)^2\,V\,, \qquad
 C_{vv|\Delta V} = \frac{1 - (S/s)^3}{s}\,,\nonumber\\
 C_{\delta\delta|\Delta V} &= s(1 - S/s)^3\,, \nonumber\\
 C_{\delta v|\Delta V} &= \frac{(1 - S/s)^2}{2} (1 + 2S/s)\,,
\end{align}
and 
\begin{align}
  \avg{\!\delta|v,\Delta,V\!} &= \Delta + S \psi V + \frac{C_{\delta v|\Delta V}}{C_{vv|\Delta V}}\, (v - \avg{\!v|V\!}) \,,\nonumber\\
  C_{\delta\delta|v\Delta V} &= s\, \frac{3(1-S/s)^4}{4(1-S^3/s^3)}\,, \nonumber\\
 \avg{\!v|\delta,\Delta,V\!} &= 
       \frac{(\delta-\Delta)}{2s}\frac{1+2S/s}{1-S/s} - \frac{S/s}{2}\, V \,,
       \nonumber\\
 C_{vv|\delta\Delta V} &= (3/4)\,(1 - S/s)/s\,.
\end{align}

In general, since $S\psi\, V\ge 0$, the dependence on $V$ always acts to increase the effective value of $B$ in $p(b|B,V)$.  Hence, if we define a characteristic scale as that $s$ where 
 $b - B - S(1 - S/s)\,V = \sqrt{s}(1 - S/s)^{3/2}$, 
then larger $V$ means smaller $s$.  That is, walks which upcross $B$ with steeper slopes are associated with more massive progenitors at the earlier time when the barrier height was $b$.  Since steeper slopes are also associated with smaller bias factors \citep[see analysis in][and Figure~\ref{fig:ab} here for an illustrative example]{mps12}, the analysis above allows one to quantify the assembly bias effect in this model.

\subsection{Merger trees}\label{trees}

Our \eqn{condprob} is a key feature of Markov Velocity models -- one that is potentially extremely useful for fast generation of merger history trees of what is sometimes called the `main progenitor' \citep{lc93}.  This is because, in the Markov heights case, one makes independent picks from $\prod_i f(\delta_i|\delta_{i-1})$, where the $\delta_i$ need not be associated with closely spaced scales.  This boils down to making independent picks from $\prod_i p(\delta_i|\delta_{i-1})$ for each $i$ and then making a change of variables.  

In principle, our Markov Velocity models allow us to work with 
 $\prod_i f(\delta_i,v_i|\delta_{i-1},v_{i-1})$.  
In the approximation where one replaces $f(\delta_i,v_i|\delta_{i-1},v_{i-1})$ with $f_\MS(\delta_i,v_i|\delta_{i-1},v_{i-1})$, we can view $f_\MS$ of \eqn{fupBV} as an integral over terms of the form $f_\MS(\delta,v|\Delta,V)$.  The appearance of $v$ and $V$ mean that such an approach will return merger histories with more information than the traditional ones; $v$ and $V$ are expected to encode information about the halo concentration and large scale environment, so our approach leads naturally to merger histories which incorporate a form of assembly bias (c.f. Section~\ref{abias}).  We are in the process of determining if the assembly bias associated with these Markov Velocity trees is realistic.

Finally, before closing, we note that the closely related quantity, $\prod_i p(\delta_i,v_i|\delta_{i-1},v_{i-1})$ with $\delta_i=\delta_{i-1}$, also plays a key role in equation~(A3) of \citet{ms13a}, the formal exact expression for $f(s)$ (where scales $i$ and $i-1$ may be rather far apart).  Since each $p(\delta_i,v_i|\delta_{i-1},v_{i-1}) = p(v_i|v_{i-1})\,p(\delta_i|v_i,\delta_{i-1},v_{i-1})$ the product may simplify, so we are in the process of checking if the analysis of the previous section allows a fully analytic solution of the formal expression for $f(s)$ for some if not all Markov Velocity models.

\section{Discussion}

Previous work on the first crossing distribution has shown the power of including the constraint that walks must cross upwards \citep{bcek91,ms12,ms13a,ms13b}.  In particular, this has shown that studying the velocity structure of the walks -- the continuum limit of the steps -- is particularly fruitful when the steps are correlated.  
To explore this structure further, we first developed a toy model in which the correlation matrix of the steps is particularly simple (equation~\ref{corrmat}), and yet non-diagonal:  even though this matrix has no zero entries, for this process the conditional distribution of any step depends only on the one just before it (equation~\ref{pwalk}).  Therefore, this toy model exhibits the simplest level of complication one could have added to walks with completely uncorrelated steps.  

We showed how to make fast Monte Carlo realizations of such walks, providing an explicit expression for how one should think of the smoothing filter associated with the model (equation~\ref{deltaMC}).  We then used the Monte Carlos to obtain the first crossing distribution associated with the toy model, showing that it was rather similar to that for TopHat-smoothed $\Lambda$CDM walks (Figure~\ref{fig:ftoy}).  Along the way, we also used the toy model to illustrate how the distribution of walk heights -- if it is known that the barrier was crossed on a larger smoothing scale -- is modified by the correlations between steps (Figure~\ref{fig:pcross}).  

We then showed that the toy model was a special case of a more general Markov Velocities model, in which it is not the heights of the walk, but the steps, which are Markovian.  The Markov assumption allowed us to include the effects of correlations between steps rather efficiently.  We did so by first writing down the Langevin equation which governs the process (equation~\ref{langevinVsq}), showing explicitly how the scale dependence of the velocity variance $\avg{\!v^2(s)\!}$ determines the process.  We then solved the Langevin equation (equations~\ref{genMC}--\ref{Wpsi}), and studied the special case in which $\avg{\!v^2(s)\!}\propto 1/s$, arguing that such walks should be thought of as a family of scale-invariant models indexed by the constant of proportionality.  The associated first crossing distributions interpolate smoothly between the case of walks with completely correlated and completely uncorrelated steps (Figure~\ref{fig:manyg}).  Moreover, although Gaussian smoothing of scale-free power spectra produces walks that do not have Markov Velocities, their first crossing distributions are rather well approximated by those of scale-invariant Markov Velocity models having the same velocity variance structure (Figure~\ref{fig:manyg}).  

In the more general case, $\avg{\!v^2(s)\!}$ may be a more complicated function of $s$; e.g., for $\Lambda$CDM $P(k)$, $\gamma(s) = (4s\avg{\!v^2\!})^{-1/2}$ is quite well approximated by equation~(\ref{gsLCDM}).  We again provided explicit expressions for the effective smoothing filter of the white noise which affects the Langevin trajectories, arguing that equation~(\ref{psiLCDM}) should provide a good approximation for $\Lambda$CDM-like $P(k)$.  And we described our Markov Velocity Monte Carlo algorithm (equation~\ref{deltaLCDM}) for generating Markov Velocity walks whose first crossing distribution closely approximates that of TopHat-smoothed $\Lambda$CDM (Figure~\ref{fig:markov2lcdm}).  

Truncated Fourier smoothing kernels of the form given by our equation~(\ref{Wtruncated}) will generically yield Markov velocity models whatever the underlying power spectrum (Section~\ref{kmodes}).  For power-law $P(k)$, we provided an explicit mapping between the index $\gamma$ of the Markov Velocity model, the shape of the smoothing filter, and the slope of the power-law $P(k)$ (equation~\ref{an2gamma}).  For the $\Lambda$CDM family, this mapping may be combined with equation~(\ref{gsLCDM}) to get a feel for the smoothing window shape which will yield Markovian Velocities.  

In this context, TopHat smoothing of $P(k)\propto k^{-n}$ with $n=-2$ is special:  it has the same correlation structure as the Markov Velocity model obtained by smoothing with $W = 1 - (kR)^2$ (for $kR\le 1$).  Since $n=-2$ is close to the effective spectral index of the $\Lambda$CDM family of power spectra on the scales where $s\sim \delta_c^2$, this correspondence may provide an easy way to think of issues such as assembly bias \citep{st04}.  This is because, in Markov Velocity models, if the walk height and slope are known on one scale, say $S$, the walk height on smaller scales (where $s\ge S$) depends only these two values, and not on the walk heights on scales larger than $S$ (equation~\ref{condprob}).  Therefore, at fixed mass, the assembly history in Markov Velocity models should be correlated with the larger scale environment because of the dependence on slope:  assembly bias is present (equations~\ref{fupBV} and~\ref{pdvD} and related discussion, as well as Figure~\ref{fig:ab}).  However, at fixed mass {\em and} slope, there should be no further correlation between the formation history of a halo and its environment (equations~\ref{pdvDV} and~\ref{pD0DV}).  The slope of the walk associated with a protohalo patch is an indicator of the concentration of the final halo \cite[e.g.][]{dwbs08}; therefore, in Markov velocity models, there should be no correlations between formation history and environment if done at fixed halo mass {\em and} concentration.  In this sense, assembly bias effects in Markov Velocity models are relatively simple.  

The exact Markov Velocity nature of TopHat-smoothed $P(k)\propto k^{-2}$, and the fact that Markov Velocity smoothing of $\Lambda$CDM $P(k)$ yields a first crossing distribution that is in good agreement with that of TopHat smoothed $\Lambda$CDM (Figure~\ref{fig:markov2lcdm}) strongly suggest that Markov velocities are a useful approximation for future excursion set studies.  However, the first crossing distribution (of a constant barrier) for TopHat smoothed $\Lambda$CDM walks is not in as good agreement with the actual mass fractions measured in numerical simulations of halo formation.  These are rather well matched by the first crossing distribution associated with our simplest model (having $\gamma=1/2$) which provides a good description of the mass fraction in haloes (Figure~\ref{fig:comparest}).  Although one could explore how allowing the barrier height to depend on $s$ might improve the agreement, we expect even this simplest constant barrier model to provide the basis of fast Monte Carlo merger history tree algorithms which include some of the assembly effects associated with correlated steps (Sections~\ref{abias} and~\ref{trees}), and so represent a significant improvement on what is currently available.   

\section*{Acknowledgements}
The work of MM is supported by the ESA Belgian Federal PRODEX Grant no.~4000103071 and the Wallonia-Brussels Federation grant ARC no.~11/15-040.  RKS is supported in part by NASA NNX11A125G.

\bibliography{mybib}{}

\appendix

\section{Distributions after first crossing} 
In this appendix we use the Monte Carlo algorithm described in the main text to illustrate how correlations between steps modify one of the key quantities in the usual excursion set approach:  The probability $p_{\times\!}(\delta)$ that a walk reaches $\delta$ at scale $s$ having crossed the barrier at least once at larger scale.  This can be written, exactly, as
\begin{equation}
 \label{pcross}
  p_{\times\!}(\delta,s) = \!\int_0^s\!\!\dd S \!\int_{B'}^\infty \!\!\! \dd V(V-B') \,
  p(\delta, B,V, \mathrm{first~} S)\,,
\end{equation}
from which one can obtain the first crossing rate as 
\begin{equation}
 \label{otherrate}
  f(s) = -\frac{\dd}{\dd s} \int_{-\infty}^{b(s)}\!\!\!\!
  \dd \delta \Big[p(\delta,s) -p_{\times\!}(\delta,s) \Big]\,.
\end{equation}
For barriers of constant height, \eqn{otherrate} is the generalization to correlated steps of the symmetry argument used by \cite{bcek91} for walks with uncorrelated steps (in which case $p_{\times\!}(\delta)$ is a Gaussian with mean $2b$ and variance $S$).  

\begin{figure}
 \centering
 \includegraphics[width=\hsize]{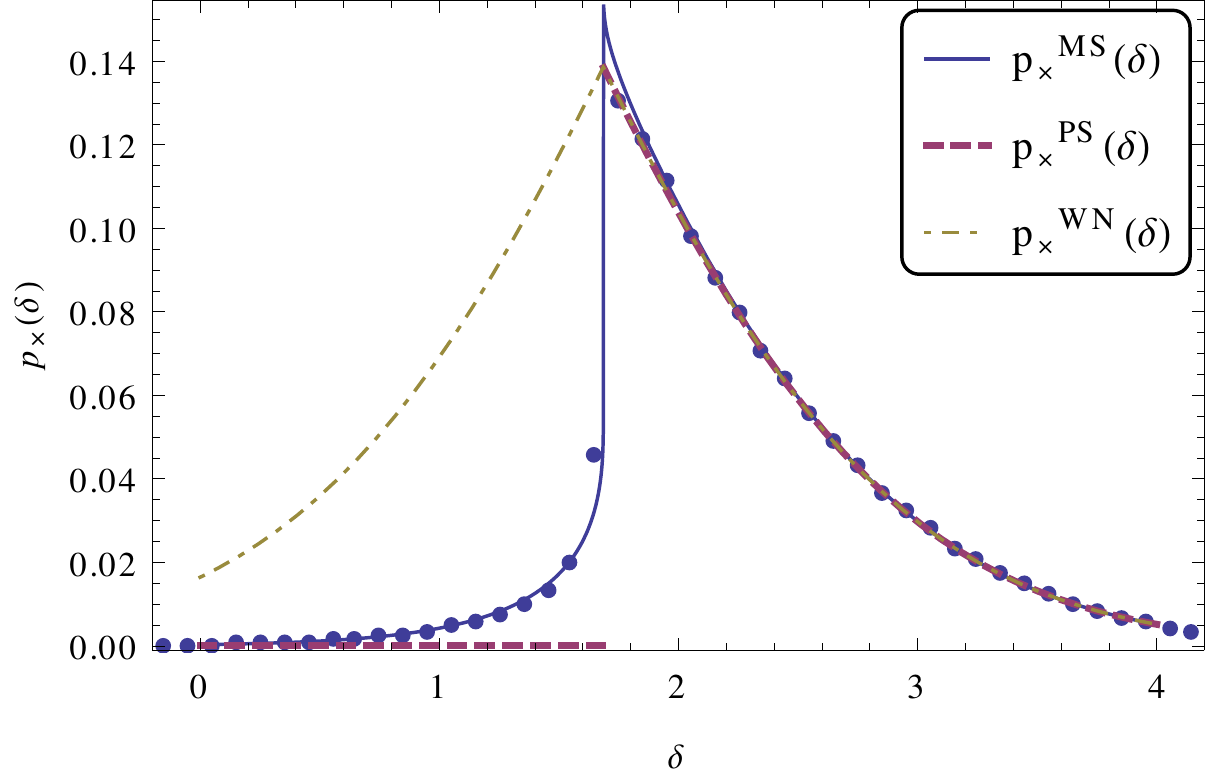}
 \caption{\label{fig:pcross}
 Probability distribution $p_\times(\delta)$ of the height $\delta$ on scale $s=2$, of walks that crossed a constant barrier of height $\delta_c=1.69$ at some $S<s$.  Filled circles show this quantity measured in Monte Carlo simulations of walks having $\avg{\!v_i,v_j\!}=s_i/s_j^2$ for $s_i<s_j$, and solid line shows \eqn{pcrosstoy}. 
For comparison, the dashed line shows $p(\delta)\vartheta(\delta-\delta_c)$, which is exact for $\delta>\delta_c$ and corresponds to walks with completely correlated steps, whereas the dot-dashed line is for walks with uncorrelated steps.}
\end{figure}

We can approximate $p_{\times\!}(\delta,s)$ by dropping from the term in the integrand of \eqn{pcross} the requirement that the walk never crossed before $S$ (this is consistent with the upcrossing approximation, $f\approx f_\MS$, of equation~\ref{fms}). We will call this approximation $p_{\times\!}^{\mathrm{MS}}(\delta,s)$.  Now, in our toy model, $p(B,V,\delta)=p(B,V)\,p(\delta-B-\psi V)$ so 
\begin{align}
 \label{pcrosstoy}
  p_\times^{\mathrm{MS}}(\delta) &=
  \int_0^s\!\frac{\dd S}{S}  \frac{2s}{s-S}
  \int_0^\infty \!\!\!\dd x \, x \,
  \frac{{\rm e}^{-x^2/2S}}{\sqrt{2\pi S}}
  \frac{{\rm e}^{-(x-2B)^2/6S}}{\sqrt{6\pi S}} \notag \\
  &\qquad\qquad \times
   \frac{{\rm e}^{-[(\delta-B)s/(s-S)-x]^2/2(s-S)}}{\sqrt{2\pi(s-S)}} \,,
\end{align}
where we have defined $x\equiv S V$.  Figure~\ref{fig:pcross} shows that this yields an excellent approximation to the Monte Carlo'd result.  It also shows the special cases in which steps are completely correlated (a step function) or completely uncorrelated (the `mirror image' of the tail of an error function).

Inserting $p_\times^{\mathrm{MS}}$ of \eqn{pcrosstoy} in place of $p_{\times\!}$ in \eqn{otherrate} gives
\begin{align}
  f(s) \simeq f_\mathrm{PS}(s) +
  \int_0^s\!\!\dd S \!\int_0^\infty \!\!\!\! \dd V \,V 
  \frac{\dd p(\delta\!\leq b, B, V+B') }{\dd s}\,,
\label{eq:correction}
\end{align}
where we have redefined $V\to V+B'$, and brought $\dd/\dd s$ inside the integral over $S$ (since its action on $s$ in the integration limit gives zero). 
For our toy model, $p(\delta\!\leq b, B, V+B')$ is known exactly, so 
\begin{align}
  sf(s) &\approx sf_\PS(s) + \int_0^1\frac{\dd y}{y} \frac{1}{1-y}
  \int_0^\infty \!\!\!\dd w \, w^2 \, \notag \\
  &\times  \frac{e^{-w^2/2y}}{\sqrt{2\pi y}}
  \frac{e^{-[w-2(b/\sqrt{s})]^2/6y}}{\sqrt{6\pi y}}
  \frac{e^{-w^2/2(1-y)}}{\sqrt{2\pi(1-y)}} .
\label{ftoy}
\end{align}
Figure~\ref{fig:ftoy} shows that this approximation for $f(s)$ is substantially more accurate than $f_\MS(s)$ of \eqn{fms}; see \cite{ms13b} for why it is not quite as accurate as $f_\mathrm{BS}(s)$.

\label{lastpage}

\end{document}